\documentclass[pra,aps,reprint,amsmath,amssymb,showkeys,showpacs]{revtex4-2}
\usepackage{graphicx,bm,mathpazo,physics,color,hyperref}
\hypersetup{colorlinks=true,linkcolor=blue,citecolor=blue,urlcolor=blue}
\begin{document}

\title{Attosecond Ionization Time Delays in Strong-Field Physics}

\author{Yongzhe~Ma}
\affiliation{State Key Laboratory of Precision Spectroscopy, East China Normal University, Shanghai 200241, China}
\author{Hongcheng~Ni}
\email{hcni@lps.ecnu.edu.cn}
\affiliation{State Key Laboratory of Precision Spectroscopy, East China Normal University, Shanghai 200241, China}
\author{Jian~Wu}
\email{jwu@phy.ecnu.edu.cn}
\affiliation{State Key Laboratory of Precision Spectroscopy, East China Normal University, Shanghai 200241, China}


\begin{abstract}

Electronic processes within atoms and molecules reside on the timescale of attoseconds. Recent advances in the laser-based pump-probe interrogation techniques have made possible the temporal resolution of ultrafast electronic processes on the attosecond timescale, including photoionization and tunneling ionization. These interrogation techniques include the attosecond streak camera, the reconstruction of attosecond beating by interference of two-photon transitions, and the attoclock. While the former two are usually employed to study photoionization processes, the latter is typically used to investigate tunneling ionization. In this Topical Review, we briefly overview these timing techniques towards an attosecond temporal resolution of ionization processes in atoms and molecules under intense laser fields. In particular, we review the backpropagation method, which is a novel hybrid quantum-classical approach towards full characterization of tunneling ionization dynamics. Continued advances in the interrogation techniques promise to pave the pathway towards the exploration of ever faster dynamical processes on an ever shorter timescale.


\end{abstract}

\maketitle

\section{Introduction}
\label{sec:introduction}

All dynamical processes have their characteristic timescales. For an isolated molecule, the motion of its atomic constituent falls on the picosecond (1 ps $=10^{-12}$ s) level for rotational dynamics, while the vibrational dynamics resides on the shorter timescale of femtoseconds (1 fs $=10^{-15}$ s). An electron within an atom, which is a more microscopic object in the hierarchy of matters compared to atoms within a molecule, possesses a faster timescale of attoseconds (1 as $=10^{-18}$ s). Remarkably, the 2023 Nobel Prize for Physics has been awarded to Prof.\ Agostini, Prof.\ Krausz, and Prof.\ L'Huillier ``for experimental methods that generate attosecond pulses of light for the study of electron dynamics in matter'', underscoring the importance of exploring ultrafast electron dynamics on its intrinsic attosecond timescale. The characteristic timescale of a dynamical process originates from the energy level spacing accessible to this process. We take an electronic process as an example. For any eigenstate of the electron, it is no doubt static. A dynamical electron wave packet can however be formed if there are two (or more) eigenstates participating in the process, which can be written as
\begin{equation}
\psi=e^{-iE_1t}\psi_1+e^{-iE_2t}\psi_2=e^{-iE_1t}(\psi_1+e^{-i\Delta Et}\psi_2),
\end{equation}
i.e., a superposition of state $\psi_1$ with energy $E_1$ and state $\psi_2$ with energy $E_2$, where $\Delta E=E_2-E_1$ is the energy difference between the two eigenstates. Clearly, the electronic wave packet has an oscillation time period of $T=2\pi/\Delta E$, with a timescale dictated by the energy level spacing. This energy separation is, in turn, determined by the particle mass and the spatial extent confining the motion of the dynamical process, exactly the reason why electronic processes are faster than atomic processes.

Typically, to time resolve a dynamical process of a certain timescale, a probe on the same or faster timescale is necessary. When a laser pulse is used as the probe, the pulse duration should generally be short compared to the timescale of the dynamical process to be measured. In the 1980s, femtosecond lasers started to develop, enabling the production of the movie of a chemical reaction \cite{Zewail2000}. Starting in the 2000s, laser sources with attosecond duration have matured, making possible the observation and control of electronic processes \cite{Krausz2009} and the formation and breakage of chemical bonds \cite{Li2022}, which are the fastest resolvable physical processes up to date.

With ever shorter pulse durations, the peak pulse intensity correspondingly increases with a certain pulse energy per laser shot. In particular, the laser intensity can be routinely amplified through the chirped-pulse amplification \cite{Strickland1985} with a table-top setup, where the laser pulse is first stretched, then amplified, and finally compressed to reach a peak intensity comparable to the Coulomb binding experienced by an electron within an atom. With such a high laser intensity, a variety of novel physical phenomena are discovered or realized, including above-threshold ionization \cite{Agostini1979,Kruit1983,Eberly1991,Becker2002}, tunneling ionization \cite{Keldysh1965,Chin1985,Popov2004,Popruzhenko2014,Amini2019}, nonsequential double ionization \cite{LHuillier1982,Fittinghoff1992,Walker1994}, and high-order harmonic generation \cite{Krause1992,Macklin1993,Popmintchev2010}, the latter of which is the base of the production of attosecond laser pulses in the extreme ultraviolet (XUV) spectral regime.

Electronic ionization under intense laser fields can generally be categorized into multiphoton ionization and tunneling ionization, the two cornerstones of strong-field physics. In this Topical Review, we briefly overview the historical debates and recent advances over the study of attosecond ionization time delays under intense laser field. The photoionization time delay, corresponding to the timescale of multiphoton ionization, as well as the tunneling time delay, corresponding to the timescale of tunneling ionization, will be covered. This Topic Review is by no means exhaustive, but rather, touches more upon the fundamental interpretations of the attosecond ionization delays in strong-field physics.

This Review is organized as follows. In Sec.~\ref{sec:ionization}, we provide a sketch of the two complementary ionization mechanisms of multiphoton ionization and tunneling ionization. In Sec.~\ref{sec:techniques}, we detail the interrogation techniques towards the time resolution of multiphoton ionization and tunneling ionization, including the attosecond streak camera \cite{Hentschel2001,Itatani2002,Pazourek2015}, the reconstruction of attosecond beating by interference of two-photon transitions (RABBITT) \cite{Paul2001,Dahlstroem2012,Dahlstroem2013}, and the attoclock \cite{Eckle2008}. In Sec.~\ref{sec:photoionization}, we overview the study of photoionization time delay. In Sec.~\ref{sec:tunneling}, we cover the topics related to tunneling time delay. A summary and perspective is given in Sec.~\ref{sec:summary}. Atomic units are used throughout unless stated otherwise.


\section{Ionization under intense laser fields}
\label{sec:ionization}

Light can be considered either in a particle perspective, i.e., as composed of photons, or in a field perspective, namely, as a classical electromagnetic field. Correspondingly, strong-field ionization can be depicted in the photon perspective or the field perspective. In the photon perspective, ionization is considered in the frequency domain, concerning the frequency spectra of the laser pulse.

\begin{figure}[b]
  \centering
  \includegraphics[width=\columnwidth]{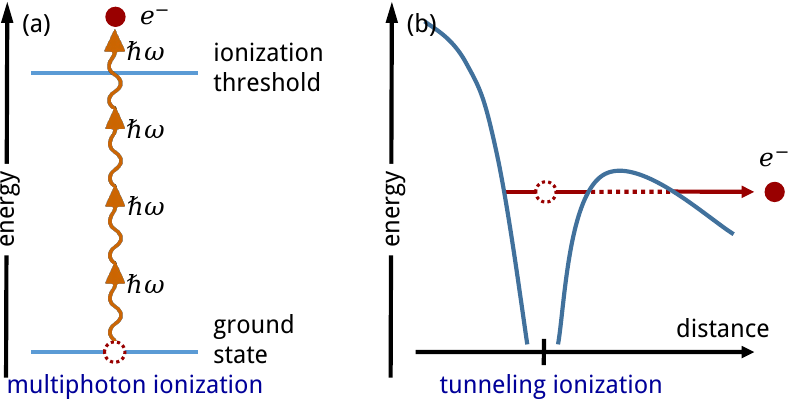}
  \caption{Two major complementary ionization mechanisms under intense laser fields. In multiphoton ionization [panel (a)], the light pulse is considered as photons in the frequency domain; while in the tunneling ionization [panel (b)], the light pulse is regarded as a classical field in the time domain.}
  \label{fig:ionization}
\end{figure}

Within an intense 800 nm laser field at an intensity of $10^{15}$ W/cm$^2$, which is routinely achievable nowadays, the cubic box of a size spanning a wavelength contains an abundant amount of coherent photons at a massive number of $10^9$. Under these conditions, ionization is no longer perturbative as suggested by photoionization. Rather, more than one photon can be absorbed collectively to overcome the ionization potential $I_p$ \cite{Goeppert1931}. Further, more photons than enough to overcome $I_p$ can even be absorbed simultaneously, thereby producing above-threshold ionization. One may refer to this class of ionization as the multiphoton ionization [Fig.~\ref{fig:ionization}(a)].

According to Bohr's correspondence principle, with such high photon density, the ionization can be described alternatively in the field perspective, namely via interaction with a classical field, concerning the temporal structure of the light pulse in the time domain while ignoring the photonic and quantum nature of photons. At a certain instance of time, the classical electric field suppresses the Coulomb potential from a certain direction, resulting in the formation of a barrier from which the bound electron may tunnel out, leading to tunneling ionization [Fig.~\ref{fig:ionization}(b)].

These two ionization mechanisms can generally be distinguished with the Keldysh parameter $\gamma$, defined as $\gamma=\frac{\omega}{F}\sqrt{2I_p}$, where $\omega$ is the angular frequency of the laser pulse and $F$ is the magnitude of the laser electric field. When $\gamma\gg1$, the ionization is in the multiphoton regime, since a large $\gamma$ indicates a relative low field magnitude $F$, i.e., small photon number, and a large photon energy $\omega$. In this case, a small number of photons participate in the ionization process, corresponding to multiphoton ionization. Conversely, when $\gamma\ll1$, the ionization is in the tunneling domain, since a small $\gamma$ implies a large field amplitude $F$, or a large number of photons, and a small photon energy $\omega$. In this case, a large number of photons participate, when the ionization is more easily depicted in the field perspective, corresponding to tunneling ionization. When $\gamma \sim 1$, which is typical for many current intense-field experiments, the ionization is in the nonadiabatic tunneling regime. In this scenario, an electron tunnels through a time-dependent potential barrier, and the energy of the electron changes during the tunneling process. This nonadiabatic effect will significantly alter the dynamic process of ionization. For example, electrons would tunnel out at a position closer to the nucleus than in the adiabatic scenario \cite{Ni2018a,Ni2018b}, and they would possess nonzero initial transverse momentum at the tunnel exit under circularly or elliptically polarized laser pulses \cite{Kaushal2013,Boge2013,Geng2014,Luo2021}, which has been directly verified by experiments \cite{Eckart2018}. Moreover, current-carrying orbitals, such as atomic $p_\pm$ orbitals, manifest different nonadiabaticities under circularly polarized laser pulses \cite{Barth2011,Barth2013,Barth2014,Liu2018}. In addition, electrons also experience nonadiabatic subcycle dynamics in tailor laser fields, such as orthogonally polarized two-color laser fields \cite{Geng2015}.

\section{Interrogation techniques}
\label{sec:techniques}

In this section, we briefly summarize the interrogation techniques for timing photoionization and tunneling ionization, including the attosecond streak camera, the RABBITT technique, and the attoclock. The former two are usually used for time resolving photoionization while the attoclock is typically employed for timing tunneling ionization.

\begin{figure*}
  \centering
  \includegraphics[width=\textwidth]{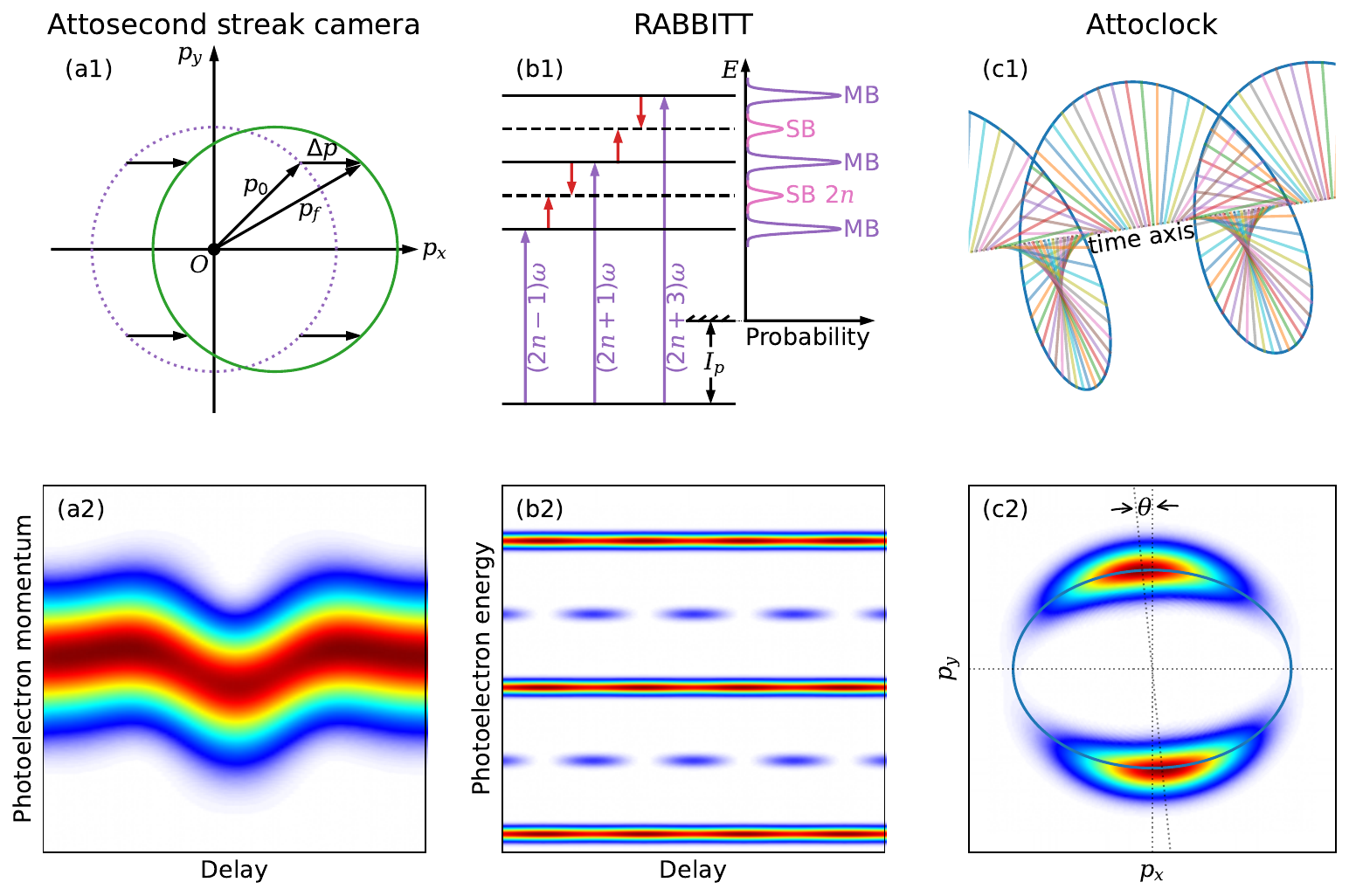}
  \caption{Interrogation techniques (first row) and their corresponding signals (second row) for attosecond electronic processes of ionization under intense laser fields. These techniques include the attosecond streak camera [column (a)], RABBITT [column (b)], and the attoclock [column (c)].}
  \label{fig:techniques}
\end{figure*}

\subsection{Attosecond streak camera}

The attosecond streak camera has evolved into an important tool for timing photoionization. It uses an isolated attosecond pulse (IAP) in the XUV spectral regime as the pump to photoionize an electron from the target and utilizes a short laser pulse typically in the infrared (IR) regime as the probe to steer the energy of the emitted electron depending on its release time, thereby mapping time onto energy (classical energy streaking). The XUV pump and IR probe pulses are typically linearly polarized along the same direction. Practically, the delay $\tau$ between the XUV and IR fields are scanned to construct a streaking trace, from which the photoionization time delay is extracted.

As shown in Fig.~\ref{fig:techniques}(a1), upon photoionization, the resulting photoelectrons are distributed along the magenta dashed circle with radius $p_0$ in the absence of the IR streaking field. When the IR streaking field is superimposed, the circle, shown as the green solid line, is displaced with respect to the IR-free case with a displacement $\Delta p$ depending on the XUV-IR delay $\tau$ along the polarization direction of the XUV and IR pulses.

The full laser field, composed of an XUV pump pulse and an IR probe pulse with a relative delay $\tau$, can be written as
\begin{align}
\bm{A}(t)&=\bm{A}_\mathrm{XUV}(t-\tau)+\bm{A}_\mathrm{IR}(t), \\
\bm{F}(t)&=\bm{F}_\mathrm{XUV}(t-\tau)+\bm{F}_\mathrm{IR}(t),
\end{align}
where $\bm{A}$ and $\bm{F}$ stand for the laser vector potential and electric field, respectively. In this Review, we use the convention that the IR probe pulse is fixed while the center of the XUV pump pulse varies in time, with a delay $\tau$ with respect to the center of the IR pulse. Note that the physical observables do not depend on the convention chosen. Now we consider the streaking scenario scanning the XUV-IR delay $\tau$. At a certain delay $\tau$, upon instantaneous photoionization initiated by the XUV pulse, the photoelectron propagates in the Coulomb potential of the parent ion within the IR streaking field, reaching a final momentum
\begin{align}
\bm{p}_f(\tau)=&\bm{p}_i(\tau)+\int_\tau^\infty\bm{a}_\mathrm{C+IR}[\bm{r}(t)]dt \nonumber \\
=&\underbrace{\bm{p}_i(\tau)+\int_\tau^\infty\bm{a}_\mathrm{C}[\bm{r}(t)]dt}_{\bm{p}_0}+\underbrace{\int_\tau^\infty\bm{a}_\mathrm{IR}[\bm{r}(t)]dt}_{-\bm{A}_\mathrm{IR}(\tau)} \nonumber \\
&+\int_\tau^\infty\{\bm{a}_\mathrm{C+IR}[\bm{r}(t)]-\bm{a}_\mathrm{C}[\bm{r}(t)]-\bm{a}_\mathrm{IR}[\bm{r}(t)]\}dt \nonumber \\
=&\bm{p}_0-\bm{A}_\mathrm{IR}(\tau)+\bm{p}_\mathrm{cont}(\tau),
\label{eq:momentum}
\end{align}
where $\bm{a}$ stands for the acceleration either under the Coulomb and laser combined field (labelled as ``C+IR''), under the Coulomb potential only (labelled as ``C''), or under the laser field only (labelled as ``IR''), $\bm{p}_i$ and $\bm{p}_f$ represent the initial and final momentum respectively, $\bm{p}_0$ is the asymptotic momentum in the absence of the IR streaking field, and $\bm{p}_\mathrm{cont}$ stands for the momentum shift accumulated in the continuum motion from potential-laser coupling. The last term $\bm{p}_\mathrm{cont}$ can be absorbed into the vector potential, resulting in the streaking trace
\begin{equation}
\bm{p}_f(\tau)=\bm{p}_0-\bm{A}_\mathrm{IR}(\tau+t_s),
\label{eq:trace}
\end{equation}
where $t_s$ is the streaking delay. This is the principle of attosecond streaking. A typical streaking trace [Eq.~\eqref{eq:trace}] is shown in Fig.~\ref{fig:techniques}(a2). At each XUV-IR delay $\tau$, the photoelectron momentum (or energy) distribution along the polarization direction of the XUV and IR pulses is recorded, which is modulated by the vector potential of the IR streaking pulse depending on the delay $\tau$, leading to the streaking trace. For each $\tau$, the expected value of the photoelectron momentum $\langle p(\tau)\rangle$ can be obtained. Fitting $\langle p(\tau)\rangle$ to Eq.~\eqref{eq:trace}, the streaking time delay $t_s$ can be obtained.

\subsection{RABBITT}

Other than the attosecond streak camera, the RABBITT technique is another important tool to time resolve ultrafast electron dynamics during photoionization. Similar to the attosecond streak camera, RABBITT also employs an XUV pump and an IR probe that are colinearly polarized for timing. In contrast to the attosecond streak camera, however, the XUV pump is not an IAP in RABBITT, but is instead an XUV attosecond pulse train (APT) produced by high-order harmonic generation (HHG) from the IR pulse. In addition, the IR probe pulse is generally weaker (typically $10^{10}<I_{\rm IR}<10^{12}$ W/cm$^2$) and longer than that used in the attosecond streak camera (typically $I_{\rm IR}\sim10^{13}$ W/cm$^2$), supporting a perturbative description of the RABBITT technique in the frequency domain.

As shown in Fig.~\ref{fig:techniques}(b1), the XUV-APT generated from rare gas atoms has a frequency spectrum with peaks at odd orders of the base frequency of the IR pulse, namely, it has a frequency of $(2n\pm1)\omega$, where $\omega$ is the frequency of the IR pulse. The photoelectrons produced by directly absorbing an XUV photon will correspondingly emit with a kinetic energy of $(2n\pm1)\omega-I_p$, which are called the main bands (MB) of the XUV-APT photoelectron spectrum. When an electron absorbs an XUV photon with frequency $(2n+1)\omega$, it could additionally emit another IR photon with frequency $\omega$, reaching an energy of $2n\omega-I_p$. Conversely, when an electron absorbs an XUV photon with frequency $(2n-1)\omega$, it could additionally absorb another IR photon with frequency $\omega$, reaching the same final energy of $2n\omega-I_p$. These two pathways would effectively interfere at the photoelectron energy of $2n\omega-I_p$, which is called the side band (SB), resulting in a modulation of the yield of the SB photoelectrons as a function of the XUV-IR delay $\tau$. A lowest-order perturbation derivation gives the probability modulation
\begin{equation}
S_{2n}=\alpha+\beta\cos\left(2\omega\tau-\Delta\phi_{2n}-\Delta\theta_{2n}\right),
\end{equation}
where $\alpha$ and $\beta$ are fitting parameters, $\tau$ is the delay of the IR pulse relative to the XUV-APT (note the different convention to that used in streaking), $\Delta\phi_{2n}$ is the phase difference between consecutive harmonics of frequencies $(2n+1)\omega$ and $(2n-1)\omega$ specific to the XUV-APT structure, and $\Delta\theta_{2n}$ is the intrinsic atomic phase related to the transition from the initial state to the $2n$ SB. From the atomic phase $\Delta\theta_{2n}$, the atomic delay $t_a$ can be defined as
\begin{equation}
t_a=\Delta\theta_{2n}/(2\omega),
\end{equation}
the counterpart quantity of the streaking delay $t_s$ as in the streaking scenario.

Although the RABBITT technique and the attosecond streak camera are conceptually distinct, in that RABBITT is the timing in the frequency domain while the attosecond streak camera is the timing applied in the temporal domain, they result in the same extracted value of the time delay, be it called the atomic delay $t_a$ or the streaking delay $t_s$ \cite{Pazourek2015}.

\subsection{Attoclock}

In contrast to the attosecond streak camera and the RABBITT technique that have been commonly used to time-resolve photoionization processes, the attoclock protocol employs a short circularly or elliptically polarized IR laser pulse as both the pump and the probe, mapping time onto the emission angle of the electron (angular streaking), as shown in Fig.~\ref{fig:techniques}(c1). Up to now, the attoclock approach has been used for timing of strong-field tunneling processes. The highly nonlinear nature of the tunneling process renders the ionization signal to be restricted near the peak of the pulse, which thus may act as a reference of ``time zero'' for the clock. In such a scenario, the IR laser pulse of the attoclock acted itself as the ionizing pulse which tunnel-ionizes the electron from the target and subsequently clocks the process by deflection of the emitted electron. Therefore, the attoclock is a self-reference technique for timing tunneling ionization.

The vector potential of the elliptically polarized laser pulse of the attoclock swipes across the polarization plane, forming the polarization ellipse, shown as the blue line in Fig.~\ref{fig:techniques}(c2), where the long axis of the ellipse is the major axis of the laser pulse while the short axis is the minor axis. The laser electric field follows the same shape. Within a certain optical cycle of the laser pulse, the laser field peaks along the major axis, where ionization probability peaks, acting thus as the ``pump'', setting ``time zero''. Hence, if the attoclock is to be considered in a pump-probe scenario, the delay between the pump and probe could be defined as $\tau=t-t_0|\max\{F(t_0)\}$, where $F(t_0)=\abs{\bm{F}(t_0)}$. The electron released from tunneling ionization is subsequently accelerated in the remaining laser field, reaching a final momentum
\begin{equation}
\bm{p}_f(\tau)=\bm{p}_i(\tau)+\int_\tau^\infty\bm{a}_\mathrm{C+IR}[\bm{r}(t)]dt,
\end{equation}
just as that in Eq.~\eqref{eq:momentum}. What's different, though, is that the tunnel exit where the classical motion starts is much farther from the nucleus where the Coulomb attraction from the parent ion is relatively small so that $\bm{a}_\mathrm{C+IR}\approx\bm{a}_\mathrm{IR}$, and the initial momentum starting from the tunnel exit is very low with $\bm{p}_i\approx\bm{0}$. Thereby, the attoclock principle follows
\begin{equation}
\bm{p}_f(\tau)\approx-\bm{A}_\mathrm{IR}(\tau).
\end{equation}

For an elliptical pulse, $t|\max\{F(t)\}=t|\min\{A(t)\}$ because the vector potential leads the electric field by $\pi/2$. In other words, electrons released along the major axis of the polarization ellipse will drift along the minor axis, provided the Coulomb potential of the parent ion can be neglected, i.e., the tick mark of ``time zero'' on the attoclock dial $\theta_0\approx\arctan[\frac{-A_{\mathrm{IR},y}(t_0)}{-A_{\mathrm{IR},x}(t_0)}]$ is along the minor axis. It follows straightforwardly the mapping between the angle and the time
\begin{equation}
\tan\theta \approx \varepsilon\tan\omega\tau,
\label{eq:mapping}
\end{equation}
where $\theta$ is the attoclock offset angle measured from $\theta_0$, as shown in Fig.~\ref{fig:techniques}(c2). We note that, usually a linear mapping between the attoclock offset angle and the time has been assumed \cite{Landsman2015,Hofmann2019,Kheifets2020,Hofmann2021}
\begin{equation}
\tau \approx \theta/\omega,
\end{equation}
which is satisfied only when $\varepsilon=1$, according to Eq.~\eqref{eq:mapping}. Needless to say, the presence of the Coulomb potential during the continuum excursion of the released electron would bend its trajectory such that the peak ionization probability may not appear near the minor axis. This Coulomb correction needs to be accounted for in order to obtain a physically meaningful time delay for the quantum tunneling process \cite{Xiao2020}. This topic will be covered later.

It is worth noting that under elliptically polarized laser pulses, the one-to-one correspondence between the photoelectron emission angle and the tunneling instant is broken near the minimum of the attoclock signal \cite{Guo2023}. This originates from the fact that two trajectories with different tunneling instants and different initial velocities could lead to the same specific final photoelectron momentum in this direction.

\section{Photoionizaton time delay}
\label{sec:photoionization}

With the interrogation techniques of the attosecond streak camera and RABBITT, time delays associated with the photoionization process can be studied. In 2007, Cavalieri \textit{et al.}\ pioneered a streaking experiment and time resolved the photoemission from the 4f core level relative to the conduction band from the tungsten surface \cite{Cavalieri2007} and reveals a relative delay of 110 as. In 2010, Schultze \textit{et al.}\ studied the photoemission from the 2s and 2p subshells of the neon atoms \cite{Schultze2010} using the attosecond streak camera and found a relative photoionization time delay of 21 as between these two subshells. Shortly after in 2011, Kl\"under \textit{et al.}\ investigated the relative photoemission time delay from the 3s and 3p subshells of the argon atoms \cite{Kluender2011} with similar conclusions employing RABBITT. These pioneering works attracted substantial research interests and had led to a plethora of following works on the topic of photoionization time delay in atoms \cite{Pazourek2015} and molecules \cite{Nisoli2017}, as well as in extended systems such as solids \cite{Siek2017,Ossiander2018} and liquids \cite{Jordan2020,Gong2022Nature,Gong2022Chimia}.

Initial wave of research works after these pioneering reports were mainly focused on the interpretation of the experimental findings, namely relative photoionization delays between different channels, when corresponding theories of photoionization time delay were developed \cite{Kheifets2010,Kheifets2013,Zhang2010,Zhang2011,Nagele2011,Pazourek2012,Nagele2012,Pazourek2013,Feist2014,Pazourek2015JPB,Su2013PRA1,Su2013PRA2,Su2013JMO,Su2014PRA,Su2014CJP,Su2014PRL,Dahlstroem2012,Dahlstroem2013}. For quite a long period of time, however, the 21 as relative delay between photoemission from 2s and 2p subshells of neon could not be explained. The most accurate theory could only account for half of the 21 as delay \cite{Pazourek2015}. Later, the discrepancy between the experiment and the theory was found to arise from the correlation-induced shake-up excitation of a second electron upon removal of the primary electron \cite{Isinger2017,Ossiander2017}. The resolution of the shake-up ionization channel requires a high frequency resolution to stand out from the direction ionization channel, which was not accomplished in earlier works.

\subsection{Breakdown of the photoionization time delay}

\begin{figure}[b]
  \centering
  \includegraphics[width=\columnwidth]{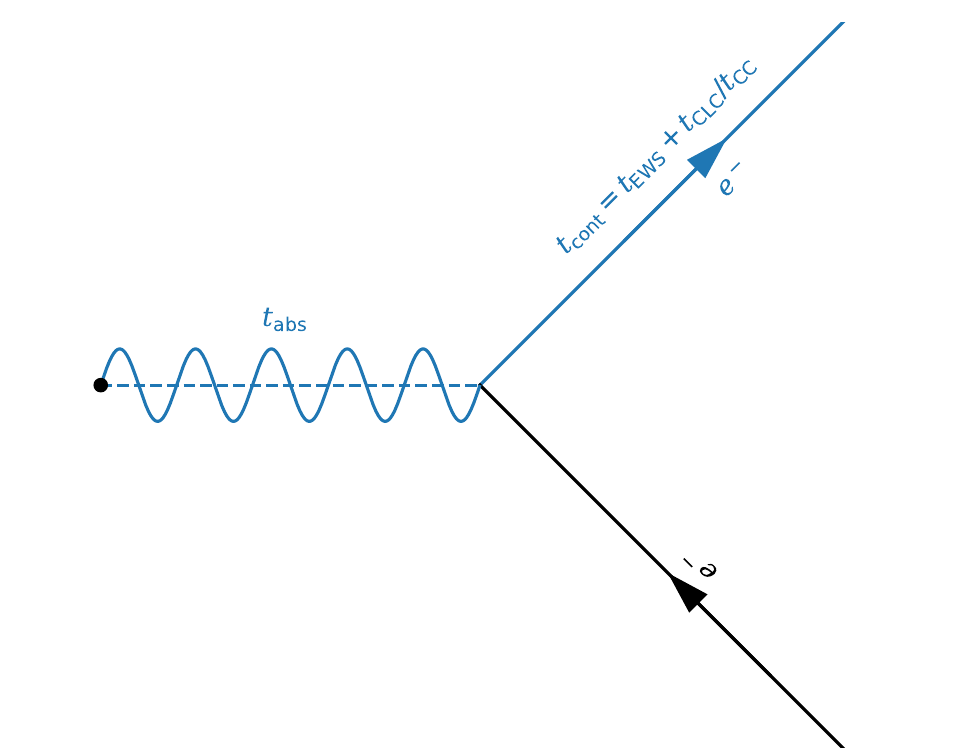}
  \caption{Diagram of the photoionization process. It could be considered as a photoabsorption process plus a half-scattering process, shown as the blue lines. The photoabsorption process is associated with a photoabsorption time delay $t_\mathrm{abs}$ and the half-scattering process comes with a continuum time delay $t_\mathrm{cont}$ that consists of a short-range EWS time delay $t_\mathrm{EWS}$ and a long-range delay induced by the measurement itself. The full scattering process is shown as solid lines.}
  \label{fig:scattering}
\end{figure}

As sketched in Fig.~\ref{fig:scattering} in blue lines, the photoionization process can generally be considered as a photoabsorption process \cite{Su2014PRL}, which promotes the bound electron into continuum, plus a half-scattering process \cite{Pazourek2015}, which brings the released electron to the remote detector. The photoabsorption process is associated with a photoabsorption time delay $t_\mathrm{abs}$ and the half-scattering process comes with a continuum time delay $t_\mathrm{cont}$ accumulated by the photoelectron in the continuum motion after phototransition, which can be further partitioned into a short-range part, the Eisenbud-Wigner-Smith (EWS) time delay $t_\mathrm{EWS}$ \cite{Eisenbud1948,Wigner1955,Smith1960}, and a long-range part. In the attosecond streak camera, the long-range part is related to the coupling of the Coulomb tail with the laser field (Coulomb-laser coupling, CLC) $t_\mathrm{CLC}$; while in RABBITT, the long-range part is considered as the delay due to continuum-continuum (CC) transition $t_\mathrm{CC}$, and can be identified as the EWS delay for CC transitions \cite{Fuchs2020}. Both the CLC delay and the CC delay arise due to the presence of the probe laser field and thus are related to the measurement process, which should be removed to single out delays associated with intrinsic target properties or the pump pulse. In the following, we will discuss how each term arises.

\subsection{Photoabsorption time delay}

The quantum transition process, initiated by photon absorption, could possess a delay which one could name the photoabsorption time delay $t_\mathrm{abs}$, as shown in Fig.~\ref{fig:scattering}. For single-photon ionization in general, the full photoionization time delay is related to the complex transition amplitude \cite{Kheifets2010,Pazourek2015}
\begin{align}
a^{1\gamma}&=-i\int_{-\infty}^\infty\langle\psi_f|\bm{r}\cdot\bm{F}_\mathrm{XUV}(t)|\psi_i\rangle e^{i(E_f-E_i)t}dt \nonumber \\
&=-i\underbrace{\langle\psi_f|\bm{r}\cdot\hat{\bm{\epsilon}}|\psi_i\rangle}_\mathrm{continuum}\underbrace{\int_{-\infty}^\infty dt F_\mathrm{XUV}(t)e^{i \Delta E_{fi}t}}_\mathrm{photoabsorption},
\end{align}
with the photoionization delay given by $d\arg\{a^{1\gamma}\}/dE$, where $\hat{\bm{\epsilon}}$ is the unit vector of laser polarization and $\Delta E_{fi}=E_f-E_i$ is the energy difference between the initial and final state.

Clearly, the full transition amplitude consists of two contributions, where the dipole $\mu_{fi}=\langle\psi_f|\bm{r}\cdot\hat{\bm{\epsilon}}|\psi_i\rangle$ is the related to the phase accumulated in the continuum motion since it contains the final scattering wave $\psi_f$, and $\int_{-\infty}^\infty dt F_\mathrm{XUV}(t)e^{i\Delta E_{fi}t}$ is associated with the photoabsorption process. The latter is the Fourier transformation of the XUV laser electric field, which, for a transform-limited laser pulse, is real. In such cases, the photoabsorption delay is zero, i.e., the photoabsorption process is considered instantaneous for single-photon ionization.

Things change, however, for two-photon ionization. Here, the full transition amplitude is written as \cite{Ishikawa2012,Su2014PRL}
\begin{align}
a^{2\gamma}&=-\sum_m\underbrace{\mu_{fm}\mu_{mi}}_\mathrm{continuum}\underbrace{\int_{-\infty}^\infty dt F_\mathrm{XUV}(t)e^{i\Delta E_{fm}t}\int_{-\infty}^t dt' F_\mathrm{XUV}(t')e^{i\Delta E_{mi}t'}}_\mathrm{photoabsorption},
\end{align}
where $m$ stands for the intermediate state. For a nonresonant two-photon ionization, where all bound intermediate states are beyond the bandwidth of the XUV pulse, the transition amplitude associated with photoabsorption reduces to
\begin{equation}
a_\mathrm{abs}^{2\gamma,\mathrm{nonres}} = \int_{-\infty}^\infty dt F_\mathrm{XUV}^2(t)e^{i \Delta E_{fi}t},
\end{equation}
which has a trivial phase, suggesting once again an instantaneous photoabsorption. On the other hand, for resonant two-photon ionization, we assume only one bound intermediate state is within the XUV bandwidth, the photoabsorption amplitude becomes
\begin{equation}
a_\mathrm{abs}^{2\gamma,\mathrm{res}} = \int_{-\infty}^\infty dt F_\mathrm{XUV}(t)e^{i\Delta E_{fm}t}\int_{-\infty}^t dt' F_\mathrm{XUV}(t')e^{i\Delta E_{mi}t'},
\end{equation}
with the corresponding photoabsorption time delay for resonant two-photon ionization
\begin{equation}
t_\mathrm{abs}^{2\gamma,\mathrm{res}}=\frac{d\arg\{a_\mathrm{abs}^{2\gamma,\mathrm{res}}\}}{dE},
\end{equation}
which is shown to be proportional to the XUV pulse duration \cite{Su2014PRL}. The existence of a photoabsorption time delay for a resonant photoionization process has been verified experimentally \cite{Gong2017}.

\subsection{Eisenbud-Wigner-Smith time delay}

\begin{figure}[b]
  \centering
  \includegraphics[width=\columnwidth]{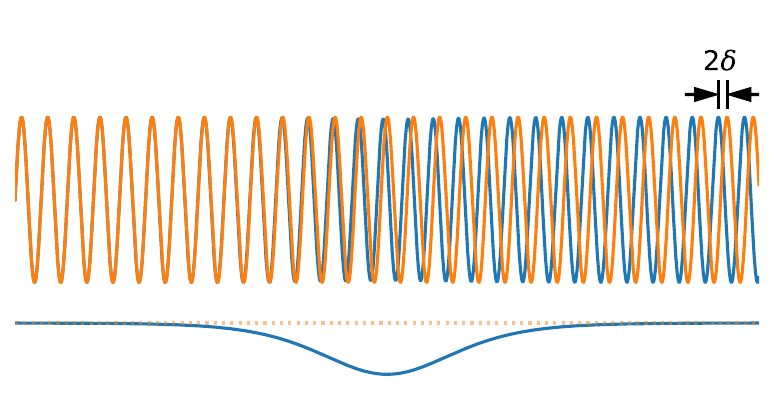}
  \caption{Illustration of potential-induced phase shift $2\delta$, giving rise to the EWS time delay $t_\mathrm{EWS}^\mathrm{scatt}=2d\delta/dE$ in the full scattering process.}
  \label{fig:ews}
\end{figure}

The EWS delay, by its original definition \cite{Eisenbud1948,Wigner1955,Smith1960}, is regarded as the time difference a particle traverses a potential as compared to its free motion. As illustrated in Fig.~\ref{fig:ews}, when an incoming wave (blue oscillatory curve) passes through the scattering region (blue dip), the wave form gets distorted, resulting in a phase shift $2\delta$ when it leaves the potential as compared to the case free of the scattering potential (orange curve). The potential effectively keeps the incoming particle to ``stick around'' for a period of time
\begin{equation}
t_\mathrm{EWS}^\mathrm{scatt}=2\frac{d\delta}{dE}
\end{equation}
in the full scattering process, where $E$ is the scattering energy of the particle. Since photoionization consists of only half of the full scattering process, the EWS time delay caused by the potential is
\begin{equation}
t_\mathrm{EWS}=\frac{d\delta}{dE}.
\end{equation}

For a long-range potential, the wave form persistently gets distorted as a particle traverses the scattering region, and thus the phase shift $\delta$ does not converge. Therefore, the EWS delay $t_\mathrm{EWS}$, by its original definition, diverges for long-range potentials, e.g., the Coulomb potential. This fact can be considered from two aspects. For one, if we carry out an asymptotic expansion of the scattering wave, the asymptotic limit of the $l$ partial wave is given by
\begin{equation}
F_l\xrightarrow{r\rightarrow\infty}\sin\left[kr-\frac{l\pi}{2}+\frac{Z}{k}\ln2kr+\sigma_l(k)\right],
\end{equation}
where $k=\sqrt{2E}$ is the scattering momentum, $Z$ is the charge of the Coulomb potential, and
\begin{equation}
\sigma_l(k)=\arg\Gamma\left(1+l-i\frac{Z}{k}\right)=\Im\left[\ln\Gamma\left(1+l-i\frac{Z}{k}\right)\right]
\end{equation}
is the Coulomb phase shift. Compared to the free wave $F_l=\sin(kr)$, the potential-induced phase shift is
\begin{equation}
\delta=-\frac{l\pi}{2}+\frac{Z}{k}\ln2kr+\sigma_l(k),
\end{equation}
which diverges for $r\rightarrow\infty$, together with the associated EWS delay.

The divergence problem is reflected in another aspect. If one trace back the evolving trajectory of the emitted electron by a straight line until the origin where the electron is assumed to be released, one may find a nonzero intercept in time, which has been interpreted as the photoionization time delay \cite{Schultze2010}. However, for a Coulomb potential, the location of the intercept depends on the remote region where one draws the straight line \cite{Su2013JMO}.

To overcome the divergence problem, one simply removes the divergent part $\frac{Z}{k}\ln2kr$ from the phase due to the long-range interaction, which arises purely because of the presence of a Coulomb tail in the potential. Thereby, one could define a short-range EWS time delay for the problem
\begin{equation}
t_\mathrm{EWS}=\frac{d\sigma_l}{dE}.
\end{equation}
Accordingly, the original definition of the EWS delay is altered such that it is not compared to the free motion, but instead, compared with respect to the motion in a potential with a Coulomb tail. The EWS time delay, thereby, gains the meaning of the time delay due to transient trapping in a short-range potential, or more precisely, in the short-range part of a potential. Clearly, in the definition of the EWS time delay, the IR probe field is not present, and thus is a specific property intrinsic to the target itself.

Now the question is, what do the attosecond streak camera and RABBITT measure? Do they measure the original EWS delay, which is divergent, or do they measure the short-range EWS delay, which stays finite? With the hindsight that the experimental results never diverge, both techniques must measure the short-range EWS delay, which we hereafter name simply as the EWS delay $t_\mathrm{EWS}$. As a matter of fact, the measured delay $t_s$ or $t_a$ does not diverge because it is related to a finite range in space that the released electron traverses during its continuum excursion until the probe pulse ceases \cite{Su2013PRA2,Su2014PRA}.

\subsection{Measurement-induced time delay}

According to Eq.~\eqref{eq:momentum}, the continuum delay $t_\mathrm{cont}$ is accumulated duration the electron excursion in the continuum after phototransition. During this process, the electron travels under the combined influence of the ionic potential and the laser field until reaching the detector. Nevertheless, a EWS time delay $t_\mathrm{EWS}$, which is dependent only on the properties of the potential itself, can be separated out. The other part of the continuum delay $t_\mathrm{cont}$ is then regarded as due to the measurement, arising from the coupled interaction of the probe laser field and the potential.

In the protocol of the attosecond streak camera, the measurement-induced delay is called the CLC delay \cite{Pazourek2015}
\begin{equation}
t_\mathrm{CLC}(E) = \frac{Z}{(2E)^{3/2}}\left[2-\ln\left(ET_\mathrm{IR}\right)\right],
\end{equation}
where $T_\mathrm{IR}=2\pi/\omega_\mathrm{IR}$ is the period of the streaking IR field.

On the other hand, in the RABBITT technique, such measurement-induced delay is named as the CC delay \cite{Dahlstroem2012}
\begin{equation}
t_\mathrm{CC}(k) = \frac{\phi_\mathrm{CC}(k,\kappa_+) - \phi_\mathrm{CC}(k,\kappa_-)}{2\omega_\mathrm{IR}},
\end{equation}
where $\kappa_\pm = k\pm\omega_\mathrm{IR}/k$, and
\begin{equation}
\phi_\mathrm{CC}(k,\kappa) = \arg\left\{\frac{(2\kappa)^{iZ/\kappa}}{(2k)^{iZ/k}}\frac{\Gamma\left[2+iZ(1/\kappa-1/k)\right]+\gamma(k,\kappa)}{(\kappa-k)^{iZ(1/\kappa-1/k)}}\right\},
\end{equation}
with
\begin{equation}
\gamma(k,\kappa) = iZ\frac{(\kappa-k)\left(\kappa^2+k^2\right)}{2\kappa^2k^2}\Gamma\left[1+iZ(1/\kappa-1/k)\right].
\end{equation}

\begin{figure}[t]
  \centering
  \includegraphics[width=\columnwidth]{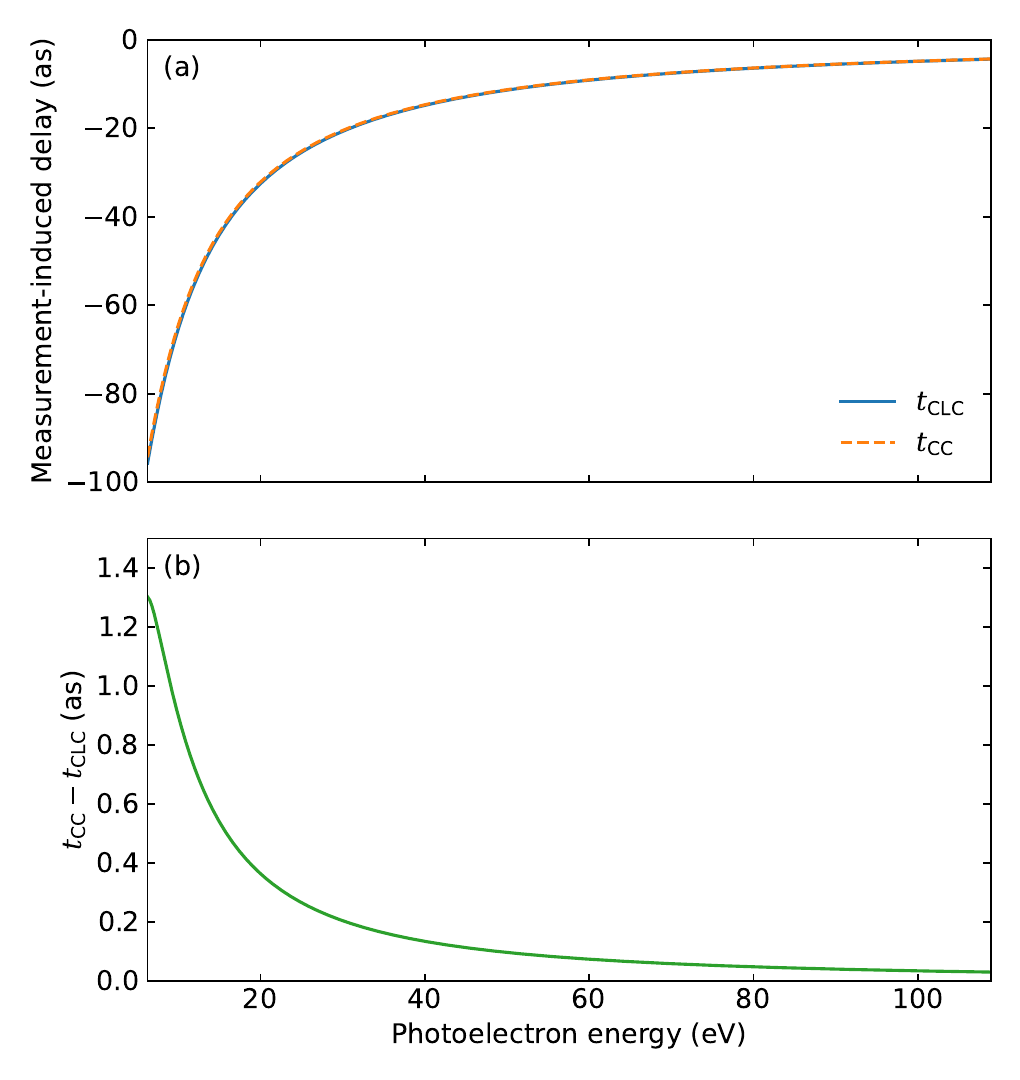}
  \caption{(a) Comparison of the CLC delay for the attosecond streak camera and the CC delay for RABBITT, whose agreement suggests the same origin of measurement nature. (b) Difference between the CC delay and the CLC delay, which is very small and generally below 1 as.}
  \label{fig:clc}
\end{figure}

A comparison of the CLC delay and the CC delay is shown in Fig.~\ref{fig:clc}. Clearly, the CLC delay and the CC delay almost coincident with each other, although they are derived in completely different ways from the time domain and the frequency domain, respectively. Indeed, they characterize the same measurement-induced contribution within the experimentally obtained streaking delay $t_s$ (for the attosecond streak camera) or atomic delay $t_a$ (for RABBITT).

Both the EWS delay and the measurement-induced delay ($t_\mathrm{CLC}$ or $t_\mathrm{CC}$) arise from the continuum motion after the electron transits to the continuum. Therefore, their contributions to the measurements can be well accounted for with classical trajectory Monte Carlo simulations \cite{Nagele2011,Su2013PRA2}.

In brief, the photoionization time delay can be measured using the attosecond streak camera and RABBITT. In the attosecond streak camera, the experimentally obtained time delay is called the streaking delay $t_s$
\begin{align}
t_s &= t_\mathrm{abs} + t_\mathrm{cont} \nonumber \\
&= t_\mathrm{abs} + t_\mathrm{EWS} + t_\mathrm{CLC}.
\end{align}
In contrast, in the RABBITT technique, the experimentally accessible time delay is named the atomic delay $t_a$
\begin{align}
t_a &= t_\mathrm{abs} + t_\mathrm{cont} \nonumber \\
&= t_\mathrm{abs} + t_\mathrm{EWS} + t_\mathrm{CC}.
\end{align}

\subsection{Recent advances}

Photoionization time delay has usually been observed along the laser polarization direction. Studying the signal along other directions, however, Heuser {\it et al.}\ \cite{Heuser2016} found an angular variation of the observed delay by a RABBITT setup. Starting from the $1s$ state of the helium atom, the electron is promoted to the $p$ continuum upon photoionization by the XUV pump pulse, with an associated EWS time delay that is expected to be isotropic. After absorbing or emitting another IR photon, however, a coherent summation of $s$ and $d$ waves results, leading to the observed angular dependence of the delay. Not surprisingly, similar situations occur in the shake-up ionization channel of the helium atom \cite{Donsa2020}. If the target is a heavier rare gas atom where the ionizing shell is $np$, the EWS time delay itself also becomes angular dependent \cite{Ivanov2017,Busto2019}. In addition, near the Fano resonance of Ar, a strong anisotropic photoionization delay has been found \cite{Cirelli2018}. Where there are more than one channels in the final continuum, how could one separate the contributions of individual partial waves? Jiang {\it et al.}\ has recently shown that it is possible by an atomic partial wave meter \cite{Jiang2022}, where the probe IR pulse is polarization-skewed \cite{Ji2019,Pan2021}, with a polarization direction rotating from parallel to perpendicular to the XUV pump pulse. This provides abundant amount of information from which the individual partial wave contributions can be singled out. The angular anisotropy of photoemission can also be exploited in such a way that timing can be achieved from a single-shot measurement via the circular holographic ionization-phase meter \cite{Donsa2019}. It also enables a reconstruction of the attosecond movie of photoemission \cite{Autuori2022}. Another associated issue of the RABBITT scheme is that the measured delay is related to the phase accumulated in the two-photon process instead of the one-photon phase reaching the SB, the timing of which is needed. This problem has been partially solved recently by a phase retrieval algorithm based on optimization \cite{Peschel2022}.

Photoionization time delay around the molecular shape resonance \cite{Huppert2016,Biswas2020,Nandi2020} has attracted considerable attention recently, especially in the context of stereo resolution in the molecular frame, which is possible from the combination of the attosecond interrogation techniques and the Cold-Target Recoil-Ion Momentum Spectroscopy (COLTRIMS) \cite{Doerner2000,Ullrich2003}. In the shape resonance, the molecular potential, together with the centrifugal potential arising from the $l$ angular momentum channel of the emitting photoelectron, forms a transient potential barrier, which traps the departing electron for a certain period of time. When the photoelectron has a certain energy close to the quasibound state of the trapping potential, a resonance results, whose resonant energy location depends on the shape of the potential, which is exactly the reason for the naming of the shaping resonance. Following the pioneering work by Vos {\it et al.}\ \cite{Vos2018}, a number of research groups have successfully resolved the stereo photoionization time delay in the molecular frame \cite{Rist2021,Heck2021,Holzmeier2021,Gong2022PRX,Ahmadi2022}, especially concerning that arising from the shape resonance. In particular, for an asymmetric molecule, the shape-resonance-induced photoionization time delay is asymmetric due to a mixture of different partial waves, or can be interpreted alternatively by the asymmetric molecular potential landscape \cite{Gong2022PRX}.

The molecular structure itself also leaves an essential footprint on the photoionization time delay. For photoemission from diatomic molecules, the constructive and destructive interference between the wave packets emitted or scattered from the two atoms leads to a modulation of the photoionization time delay depending on the photoelectron energy \cite{Ning2014,Liao2021,Heck2022}. In this scenario, the nuclear motion during bond softening also leads to a substantial effect on the photoionization time delay \cite{Wang2021}, which has been measured recently \cite{Cattaneo2022}.

Generally speaking, attosecond electron dynamics is an important topic in ultrafast science and its study has led to a variety of novel findings \cite{Kheifets2023}. Recently, the temporal resolution of the buildup of Fano resonance \cite{Gruson2016,Kaldun2016}, chiral photoemission \cite{Beaulieu2017}, Auger process \cite{Hutten2018,Li2022Science} and Rabi dynamics \cite{Nandi2022} has also been achieved. In addition, correlation effects in photoionization of argon \cite{Magrakvelidze2015} and confinement effects in photoionization of argon in C$_{60}$ \cite{Dixit2013} have been studied using the time-dependent density functional theory.

\section{Tunneling time delay}
\label{sec:tunneling}

Ever since the birth of quantum mechanics, tunneling phenomena have attracted extensive attention, in particular the question of how much time it takes to tunnel \cite{MacColl1932} has been debated frequently \cite{Landauer1994,Orlando2014}. Now that the attoclock protocol \cite{Eckle2008} has been proposed, an experimental measurement of the tunneling time delay has practically come into reach. It has thus triggered another heated wave of research over tunneling time delay. However, this topic has remained controversial \cite{Landsman2015,Hofmann2019,Kheifets2020,Hofmann2021}.

Considering the simple-man model of the attoclock
\begin{equation}
\bm{p}_f(t)=\bm{p}_i(t)+\int_t^\infty\bm{a}_\mathrm{C+IR}[\bm{r}(t')]dt' \approx-\bm{A}_\mathrm{IR}(t),
\label{eq:attoclock}
\end{equation}
it makes the assumptions that the electrons is released at rest and the effect of the Coulomb potential can be neglected. In reality, however, the combined interaction of the Coulomb and the laser fields steers the continuum excursion of the released electron such that it ends up with an additional angular offset to that induced by any possible tunneling delay. Hence, similar to the case of photoionization time delay, the measured attoclock time delay $t_\mathrm{ac}$ contains two contributions
\begin{equation}
t_\mathrm{ac} = t_\mathrm{tun} + t_\mathrm{cont},
\end{equation}
where $t_\mathrm{tun}$ is the tunneling time delay, i.e., the time the electron appears at the tunnel exit with respect to the field maximum, and $t_\mathrm{cont}$ is the delay accumulated in the continuum motion in the Coulomb and laser fields. However, $t_\mathrm{cont}$ is generally unknown and nontrivial to obtain. It has been because of this difficulty that the topic of tunneling time delay has been heavily debated over.

Basically, there are two opinions regarding tunneling time delay, one claiming none or negligible delay \cite{Eckle2008Science,Pfeiffer2012,Torlina2015,Ni2016,Eicke2018,Klaiber2018,Bray2018,Quan2019,Sainadh2019,Yu2022}, while the other claiming a finite positive delay \cite{Landsman2014,Zimmermann2016,Teeny2016,Camus2017,Douguet2018,Yuan2019,Ramos2020}. In 2008, Eckle \cite{Eckle2008Science} {\it et al.}\ employed the attoclock for the first time, placing an upper limit of 34 as and an intensity-averaged upper limit of 12 as to the tunneling time delay for the helium atom. In 2012, Pfeiffer {\it et al.}\ \cite{Pfeiffer2012} confirmed a vanishing tunneling time delay. In 2014, Landsman {\it et al.}\ \cite{Landsman2014} claimed a positive tunneling time delay, initiating the debate over the existence of a finite tunneling delay as measured with the attoclock protocol. Over time, a number of different methods have been developed to study the tunneling time delay. Below we discuss a few general and commonly used methods.

\subsection{Time-dependent Schr\"odinger equation}

Tunneling is a purely quantum mechanical process, and the description of tunneling ionization is naturally most precisely treated with direct solution of the time-dependent Schr\"odinger equation (TDSE), where the corresponding wave function evolves under the full interaction of the Coulomb and laser fields. After the laser pulse ceases, the wave function can be projected onto the Coulomb wave to extract the asymptotic momentum distribution. The numerical momentum distribution can be compared directly to the experimentally measured momentum distribution.

In 2019, Sainadh {\it et al.}\ \cite{Sainadh2019} studied the tunneling ionization of the atomic hydrogen, the simplest system where a numerical computational can be performed with no influence from electron-electron correlations. Excellent agreement between numerical and experimental offset angles was achieved. Switching to a short-range Yukawa potential, the numerical attoclock offset angle becomes zero, which provides an indirect evidence that the experimentally measured attoclock offset angle is purely due to the presence of the long-range Coulomb potential. Thereby, a zero time delay associated with the quantum tunneling process can be concluded.

Although a direct numerical solution of the TDSE provides the most accurate theoretical results, it does not come with a transparent physical picture where the dynamical evolution of the tunneling process can be separated. In particular, it remains unclear how the tunneled wave packet emerges. Other methods are thus needed for more transparent tunneling dynamics.

\subsection{Trajectory-based methods}

Evolving trajectories are the basic building blocks of classical mechanics, which provide transparent physical pictures to a dynamical process. Remarkably, the attoclock principle [Eq.~\eqref{eq:attoclock}] is a direct consequence of the physical intuition originating from the classical trajectories. Within the trajectory framework, the Coulomb interaction from the parent ion and the laser field can be considered simultaneously \cite{Xiao2016}.

The classical trajectory Monte Carlo (CTMC) method is a typical trajectory method to obtain photoelectron momentum distribution. Starting from a set of initial conditions of an electron trajectory $(t_r,\bm{r}_i,\bm{p}_i)$, including the tunneling exit time $t_r$, the tunneling exit position $\bm{r}_i$, and the initial tunneling momentum $\bm{p}_i$, the final momentum associated with the electron trajectory can be uniquely determined according to the dynamical evolution given by Eq.~\eqref{eq:attoclock}. Weighting each electron trajectory by the corresponding tunneling ionization probability, the full photoelectron momentum distribution can be constructed and compared to experimental or TDSE results.

\subsubsection{Preparing initial conditions}

There are a variety of ways to set initial conditions for the electron trajectories, which can be prepared in such a way that different levels of nonadiabatic tunneling effects are considered. Generally, the initial conditions can be prepared by the strong-field approximation (SFA) with additional saddle-point approximation (SPA) \cite{Popruzhenko2014,Amini2019}, the saddle-point approximation with nonadiabatic expansion (SPANE) \cite{Frolov2017,Ni2018b,Ni2020,Ma2021,Mao2022}, or the Ammosov-Delone-Krainov (ADK) theory \cite{Ammosov1986,Delone1998,Ivanov2005}. Among these methods, SPA includes full nonadiabatic tunneling effects, ADK comes with a simple analytical ionization rate which regards the tunneling process fully static and thus neglects any nonadiabaticity, and SPANE lies in between which accounts for most of the nonadiabaticity while retaining a simple analytical form. We note that SPANE is sometimes called the strong-field approximation with adiabatic expansion (SFAAE) \cite{Ma2021}. Below we brief these approaches to prepare initial conditions for trajectory-based simulations for an atomic target.

\textit{Preparing initial conditions using SPA.} The key idea of SFA is to neglect the Coulomb interaction in the final state and the laser interaction in the initial state, so that both states can be expressed in closed analytical forms. The complex transition amplitude to final momentum $\bm{p}$ is written as
\begin{equation}
M_{\bm{p}}^\mathrm{SFA}=-i\int_{-\infty}^{+\infty} \langle \psi_{\bm{p}} | \bm{r}\cdot\bm{F}(t) | \psi_0 \rangle dt,
\label{eq:SFA}
\end{equation}
where the final state is represented by the Volkov state (in the length gauge)
\begin{equation}
\psi_{\bm{p}}(\bm{r},t)=\exp\left\{i\left[\bm{p}+\bm{A}(t)\right]\cdot\bm{r}-\frac{i}{2}\int^t\left[\bm{p}+\bm{A}(t')\right]^2dt'\right\},
\end{equation}
which is the eigenstate of an electron in a laser field, and the initial state is
\begin{equation}
\psi_0(\bm{r},t)=\psi_0(\bm{r})e^{iI_pt},
\end{equation}
assumed to be unperturbed by the external laser field.

With a further SPA \cite{Nayak2019,Jasarevic2020}, SFA can be applied to prepare initial conditions for the electron trajectories. To this end, the transition amplitude expressed as an integral [Eq.~\eqref{eq:SFA}] is approximated by a sum over the saddle points
\begin{equation}
M_{\bm{p}}^\mathrm{SPA}=\sum_{t_s}\frac{\exp(iS_s)}{\{[\bm{p}+\bm{A}(t_s)] \cdot \bm{F}(t_s)\}^{\alpha/2}},
\label{eq:MpSPA}
\end{equation}
where the saddle-point time $t_s=t_r+it_i$ is a complex solution to the saddle-point equation (SPE)
\begin{equation}
\frac{1}{2}[\bm{p}+\bm{A}(t_s)]^2+I_p=0,
\label{eq:SPE}
\end{equation}
the phase $S_s$ is
\begin{equation}
S_s=-\int_{t_s}^{t_r} \left\{\frac{1}{2}[\bm{p}+\bm{A}(t)]^2+I_p\right\}dt,
\end{equation}
and $\alpha=1+Z/\sqrt{2I_p}$ \cite{Gribakin1997,Kjeldsen2006,Milosevic2006} with $Z$ the asymptotic charge of the parent ion.

It is generally a demanding task to solve for the saddle-point time $t_s$ from the SPE [Eq.~\eqref{eq:SPE}] directly since one need to search through the full complex domain associated with the dynamical evolution. However, if one make a specific coordinate transformation \cite{Ni2020,Ma2021,Mao2022}
\begin{equation}
(p_x,p_y,p_z)\rightarrow(t_r,k_\perp,p_z),
\end{equation}
the task becomes much simpler. Here, $t_r=\Re t_s$ is the real part of the saddle-point time $t_s$, representing the time the electron appears at the tunnel exit, and $k_\perp$ is an auxiliary momentum, where the $\perp$ subscript stands for quantities in the laser polarization plane. Making a substitution $\bm{k}_\perp = \bm{p}_\perp+{\rm Re}\bm{A}(t_s)$, the imaginary part of the SPE [Eq.~\eqref{eq:SPE}] becomes
\begin{equation}
k_x\Im A_x(t_s)+k_y\Im A_y(t_s)=0,
\end{equation}
where $x-y$ is the laser polarization plane and $z$ is the laser propagation direction. Therefore, by choosing the auxiliary momentum $k_\perp$ in the polarization plane as \cite{Ni2020,Ma2021,Mao2022}
\begin{equation}
k_\perp = \left[\bm{p}_\perp+\Re\bm{A}(t_s)\right]\cdot\frac{-\Im A_y(t_s)\hat{\bm{e}}_x + \Im A_x(t_s)\hat{\bm{e}}_y}{\Im A(t_s)},
\end{equation}
where $\Im A(t_s) = \sqrt{\left[\Im A_x(t_s)\right]^2+\left[\Im A_y(t_s)\right]^2}$,
the imaginary part of the SPE is fulfilled automatically for arbitrary pulse shapes. Notably, it applies in the nondipole scenario as well \cite{Mao2022}. The search for the saddle points is hence reduced to solving for $t_i=\Im t_s$ from the real part of the SPE
\begin{equation}
\frac{1}{2}k_\perp^2+\frac{1}{2}p_z^2-\frac{1}{2}\left[\Im A(t_r+it_i)\right]^2+I_p=0
\label{eq:ReSPE}
\end{equation}
for each point in the coordinate system $(t_r,k_\perp,p_z)$. The tunneling exit position corresponding to this saddle-point time $t_s$ is then
\begin{equation}
\bm{r}_i^\mathrm{SPA}=\Re\int_{t_s}^{t_r}[\bm{p}+\bm{A}(t)]dt=\Im\int_0^{t_i}\bm{A}(t_r+it)dt,
\label{eq:rSPA}
\end{equation}
which can be used as the initial position of the electron trajectory. The initial momentum is given by
\begin{equation}
\bm{p}_i^\mathrm{SPA}(t_r) = \bm{p} + \bm{A}(t_r),
\label{eq:pSPA}
\end{equation}
where the asymptotic momentum $\bm{p}$ is expressed in the $(t_r,k_\perp,p_z)$ coordinate as
\begin{equation}
\begin{cases}
p_x = -k_\perp\frac{\Im A_y(t_r+it_i)}{\sqrt{k_\perp^2+p_z^2+2I_p}}-\Re A_x(t_r+it_i), \\
p_y =  k_\perp\frac{\Im A_x(t_r+it_i)}{\sqrt{k_\perp^2+p_z^2+2I_p}}-\Re A_y(t_r+it_i), \\
p_z =  p_z.
\end{cases}
\end{equation}

In brief, to prepare the initial conditions for the trajectories using SPA, one may work in the coordinate system $(t_r,k_\perp,p_z)$. Aside from computational advantages, the new coordinate $(t_r,k_\perp,p_z)$ unwraps the tunneling dynamics along the time axis $t_r$ such that time-resolved tunneling dynamics is achieved automatically. Iterating through points in this coordinate system, one first obtain the imaginary part of the saddle-point time $t_i=\Im t_s$ from Eq.~\eqref{eq:ReSPE}. Subsequently, the full set of initial conditions $(t_r,\bm{r}_i,\bm{p}_i)$ are given, where the initial position is written as in Eq.~\eqref{eq:rSPA} and the initial momentum is expressed by Eq.~\eqref{eq:pSPA}. The weight of the corresponding electron trajectory is given by the squared norm of Eq.~\eqref{eq:MpSPA}. Needless to say, since a coordinate transformation has been applied, a corresponding Jacobian factor is needed to correct the weight of the electron trajectory.

\textit{Preparing initial conditions using SPANE.} Initial conditions obtained from SPA include nonadiabatic tunneling effects accurately. Although a new coordinate system can be applied to reduce the computational overhead, the calculation is still quite involved and does not give a closed analytical form of the initial conditions. In this sense, SPANE is a good alternative, which has a simple analytical form while retaining nonadiabatic tunneling effects.

To this end, we expand the vector potential to second order in $t_i$
\begin{equation}
\bm{A}(t_r+it_i) \approx \bm{A}(t_r) - it_i\bm{F}(t_r)+\frac{1}{2}t^2_i\dot{\bm{F}}(t_r).
\end{equation}
Inserting it into the SPE [Eq.~\eqref{eq:SPE}] and keeping terms up to $t_i^2$, we have
\begin{equation}
\bm{k}_\perp\cdot\bm{F}(t_r)=0,
\label{eq:imagSPANE}
\end{equation}
where the auxiliary $\bm{k}_\perp=\bm{p}_\perp+\bm{A}(t_r)$ is defined differently from that in SPA, with its value given by
\begin{equation}
k_\perp = \left[\bm{p}_\perp+\bm{A}(t_r)\right]\cdot\frac{F_y(t_r)\hat{\bm{e}}_x - F_x(t_r)\hat{\bm{e}}_y}{F(t_r)},
\end{equation}
representing the initial transverse momentum at the tunnel exit, and
\begin{equation}
t_i = \sqrt{\frac{k_\perp^2+p_z^2+2I_p}{F^2(t_r)-\bm{k}_\perp\cdot\dot{\bm{F}}(t_r)}},
\label{eq:tiSPANE}
\end{equation}
where $F(t_r)=\sqrt{F_x^2(t_r)+F_y^2(t_r)}$ is the instantaneous electric field magnitude.

The imaginary part of the phase $S_s$ can be subsequently obtained as
\begin{align}
\Im S_s & = I_pt_i + \frac{1}{2}\Re \int_{0}^{t_i}\left[\bm{p}+\bm{A}(t_r+it)\right]^2dt \nonumber \\
& \approx \left(I_p+\frac{p_z^2}{2}\right)t_i + \frac{1}{2}\Re \int_{0}^{t_i}\left[\bm{k}_\perp-it\bm{F}(t_r)+\frac{1}{2}t^2\dot{\bm{F}}(t_r)\right]^2dt  \nonumber \\
& \approx \frac{\left(k^2_\perp+p_z^2+2I_p\right)^{3/2}}{3\sqrt{F^2(t_r)-\bm{k}_\perp\cdot\dot{\bm{F}}(t_r)}}.
\end{align}
On the other hand, the real part of the phase $S_s$ is
\begin{align}
\Re S_s & = -\frac{1}{2}\Im \int_{0}^{t_i}\left[\bm{p}+\bm{A}(t_r+it)\right]^2dt \nonumber \\
& \approx -\frac{1}{2}\Im \int_{0}^{t_i}\left[\bm{k}_\perp-it\bm{F}(t_r)+\frac{1}{2}t^2\dot{\bm{F}}(t_r)\right]^2dt \nonumber \\
& \approx 0,
\end{align}
with negligible contribution to the phase of the tunneling ionization amplitude up to second order in $t_i$. In addition, the prefactor in Eq.~\eqref{eq:MpSPA} can be approximated, up to first order in $t_i$, as
\begin{align}
& \{[\bm{p}+\bm{A}(t_s)] \cdot \bm{F}(t_s)\}^{-\alpha/2} \nonumber \\
\approx & \left\{-i\sqrt{\left(k_\perp^2+p_z^2+2I_p\right)\left[F^2(t_r)-\bm{k}_\perp\cdot\dot{\bm{F}}(t_r)\right]}\right\}^{-\alpha/2}.
\end{align}
Thereby, the tunneling ionization amplitude can be expressed in the $(t_r,k_\perp,p_z)$ coordinate as
\begin{align}
M_{(t_r,k_\perp,p_z)}^\mathrm{SPANE}=&\left\{-i\sqrt{\left(k_\perp^2+p_z^2+2I_p\right)\left[F^2(t_r)-\bm{k}_\perp\cdot\dot{\bm{F}}(t_r)\right]}\right\}^{-\alpha/2} \nonumber \\
&\times \exp{-\frac{\left(k^2_\perp+p_z^2+2I_p\right)^{3/2}}{3\sqrt{F^2(t_r)-\bm{k}_\perp\cdot\dot{\bm{F}}(t_r)}}},
\end{align}
and the corresponding tunneling ionization probability as
\begin{align}
W_{(t_r,k_\perp,p_z)}^\mathrm{SPANE}=&\left\{\left(k_\perp^2+p_z^2+2I_p\right)\left[F^2(t_r)-\bm{k}_\perp\cdot\dot{\bm{F}}(t_r)\right]\right\}^{-\alpha/2} \nonumber \\
&\times \exp{-\frac{2\left(k^2_\perp+p_z^2+2I_p\right)^{3/2}}{3\sqrt{F^2(t_r)-\bm{k}_\perp\cdot\dot{\bm{F}}(t_r)}}}.
\label{eq:WSPANE}
\end{align}

With known analytical expression of $t_i$ [Eq.~\eqref{eq:tiSPANE}], the tunneling exit position can be written as
\begin{equation}
\bm{r}_i^{\mathrm{SPANE}} = -\frac{\bm{F}(t_r)}{2}\frac{k_\perp^2+p_z^2+2I_p}{F^2(t_r)-\bm{k}_\perp\cdot\dot{\bm{F}}(t_r)},
\label{eq:rSPANE}
\end{equation}
which can be used as an analytical initial position for the subsequent continuum motion. Remarkably, the tunneling exit position depends on the tunneling momentum, which has been experimentally verified recently \cite{Trabert2021}. In contrast to SPA, $k_\perp$ has a specific physical meaning in SPANE, that is, the transverse momentum perpendicular to the instantaneous polarization direction in the polarization plane, see Eq.~\eqref{eq:imagSPANE}. Therefore, at a certain moment, the direction of the electric field can be used to determine the magnitude of each component of the initial tunneling exit momentum. The corresponding initial momentum is thereby
\begin{equation}
\bm{p}_i^\mathrm{SPANE}(t_r) = \bm{p} + \bm{A}(t_r),
\label{eq:pSPANE}
\end{equation}
where the asymptotic momentum $\bm{p}$ is expressed in the $(t_r,k_\perp,p_z)$ coordinate as
\begin{equation}
\begin{cases}
p_x =  k_\perp\frac{F_y(t_r)}{F(t_r)}-A_x(t_r), \\
p_y = -k_\perp\frac{F_x(t_r)}{F(t_r)}-A_y(t_r), \\
p_z = p_z.
\end{cases}
\end{equation}

In brief, to prepare the initial conditions for the classical motion of electron trajectories using SPANE, one may as well work in the $(t_r,k_\perp,p_z)$ coordinate, just as in SPA. Iterating through this coordinate, the initial position is given by Eq.~\eqref{eq:rSPANE} with the initial momentum given by Eq.~\eqref{eq:pSPANE}. The corresponding weight for the trajectory can be assigned with the tunneling ionization rate as in Eq.~\eqref{eq:WSPANE}. It is worthwhile noting that the Jacobian factor for the coordinate transformation needs to be included in the weight as well. Remarkably, within the SPANE method, nonadiabatic effects are largely included, with nonadiabaticity only higher than order $t_i^3$ neglected. It has been shown to be able to produce an asymptotic momentum distribution similar to that given by SPA or obtained from TDSE \cite{Ma2021}.

\textit{Preparing initial conditions using ADK.} The ADK theory is widely used in studying the tunneling process, which assumes a pure static tunneling ignoring any nonadiabatic effects. The corresponding ADK tunneling rate can be retrieved from the SPANE tunneling rate [Eq.~\eqref{eq:WSPANE}] by neglecting the term leading to nonadiabaticity, i.e., by removing the term relating to the time variation in the field $\dot{\bm{F}}$, leading to
\begin{align}
W_{(t_r,k_\perp,p_z)}^\mathrm{ADK}=&\left\{\left(k_\perp^2+p_z^2+2I_p\right)F^2(t_r)\right\}^{-\alpha/2} \nonumber \\
&\times \exp{-\frac{2\left(k^2_\perp+p_z^2+2I_p\right)^{3/2}}{3F(t_r)}}.
\label{eq:WADK}
\end{align}
This ADK tunneling rate has been intentionally written in this form such that it resembles the SPANE rate [Eq.~\eqref{eq:WSPANE}]. Clearly, the only difference between them is that SPANE employs an effective field strength $\widetilde{F}(t_r)=\sqrt{F^2(t_r)-\bm{k}_\perp\cdot\dot{\bm{F}}(t_r)}$ accounting for the the field variation while the ADK rate only parametrically depends on the field strength $F(t_r)$. The ADK rate can be further expanded in terms of the relatively small transverse momentum $k_\perp$ and lateral momentum $p_z$, resulting in the more familiar form
\begin{align}
W_{(t_r,k_\perp,p_z)}^\mathrm{ADK}\approx&\left\{\left(k_\perp^2+p_z^2+2I_p\right)F^2(t_r)\right\}^{-\alpha/2} \nonumber \\
&\times \exp{-\frac{2\left(2I_p\right)^{3/2}}{3F(t_r)}} \exp{-\frac{\sqrt{2I_p}}{F(t_r)}k_\perp^2} \nonumber \\
&\times \exp{-\frac{\sqrt{2I_p}}{F(t_r)}p_z^2}.
\end{align}
In the ADK theory, therefore, both the transverse momentum $k_\perp$ and the lateral momentum $p_z$ manifest a Gaussian distribution around zero. Due to nonadiabaticity, however, $k_\perp$ comes with a shift from zero in SPA and SPANE.

The tunneling exit position of the electron is given by
\begin{equation}
\bm{r}_i^\mathrm{ADK}= -\frac{\bm{F}(t_r)}{2}\frac{k_\perp^2+p_z^2+2I_p}{F^2(t_r)},
\label{eq:rADK}
\end{equation}
which can be used as the initial position for the classical electron trajectory. Clearly, when the momentum at the tunnel exit is considered, the tunneling exit distance is no longer the commonly used value of $I_p/F(t_r)$. The corresponding initial momentum, on the other hand, has the same form as in SPANE [Eq.~\eqref{eq:pSPANE}].

Since the ADK theory is fully static, the total energy is conserved, as expected:
\begin{equation}
E^\mathrm{ADK} = \frac{k_\perp^2}{2} + \frac{p_z^2}{2} + \bm{r}_i^\mathrm{ADK}\cdot\bm{F}(t_r) = -I_p.
\end{equation}

In summary, the initial conditions for the classical propagation of the electron trajectories can be prepared using SPA, SPANE, or ADK, which account for nonadiabatic tunneling effects to different levels. In SPA, nonadiabaticity is fully included, but it does not come with a closed analytical form and its computation could be quite involved, even using the $(t_r,k_\perp,p_z)$ coordinate, since the imaginary part of the saddle-point time $t_i$ still needs to be solved from the real part of the nonlinear SPE. On the other hand, ADK has a very simple analytical form, but it is fully static, neglecting any nonadiabaticity. SPANE lies in between, which provides an analytical form without the necessity to solve for the saddle-point time while at the same time nonadiabaticity is largely taken into account of. These schemes provide the basis for using the trajectory method to study the tunneling problem, in particular, tunneling time delays using the attoclock protocol.

\subsubsection{Trajectory phase}

The CTMC method is a convenient computational framework to gain physical insights into a dynamical process. It, however, lacks any quantum element, and thus may fail when the quantum feature is prominent. For example, it fails to reproduce  fringes in above-threshold ionization arising from intercycle interference. Inspired by the Feynman's path-integral approach, Li \textit{et al.}\ \cite{Li2014} assigned a phase to each electron trajectory accumulated in the continuum motion
\begin{equation}
\Phi^\mathrm{QTMC}=-\int_{t_r}^{\infty}\left\{\frac{v^2(t)}{2}+V[r(t)]+I_p\right\}dt,
\label{eq:QTMC_phase}
\end{equation}
where $v(t)$ is the electron velocity and $V(r)$ is the Coulomb potential of the parent ion. To obtain the asymptotic momentum distribution, electron trajectories reaching the same final momentum are binned coherently, where the continuum phase [Eq.~\eqref{eq:QTMC_phase}] is included in the complex amplitude of the electron trajectories. It has been shown to play an important role in retrieving quantum features \cite{Li2014}. This scheme has been named the quantum trajectory Monte Carlo (QTMC) method \cite{Li2014}. We note that, despite the name, the trajectory itself is still purely classical and governed by Newton's equations of motion.

The semiclassical two-step model (SCTS) \cite{Shvetsov2016} improves the continuum phase given by QTMC, with
\begin{align}
\Phi^\mathrm{SCTS}=&-\bm{v}_i\cdot\bm{r}_i \nonumber \\
&-\int_{t_r}^{\infty}\left\{\frac{v^2(t)}{2}+V[r(t)]-\bm{r}(t)\cdot\bm{\nabla}V[r(t)]+I_p\right\}dt,
\label{eq:SCTS_phase}
\end{align}
where $\bm{v}_i$ and $\bm{r}_i$ are the initial momentum and position at the tunnel exit, respectively.

The continuum phases given by QTMC and SCTS are similar but differ slightly. This is because the QTMC continuum phase [Eq.~\eqref{eq:QTMC_phase}] is derived under quasi-static conditions, where the initial longitudinal velocity is zero such that the term $-\bm{v}_i\cdot\bm{r}_i$ vanishes. The extra term in the integrand of the SCTS continuum phase [Eq.~\eqref{eq:SCTS_phase}] comes from the equation of motion, which eluded QTMC while it was derived. While both the QTMC and SCTS phases work well in the attoclock scenario \cite{Ma2021}, the SCTS continuum phase has been shown to outperform that of QTMC when backscattering is prominent, especially in the case of linear polarization, giving correct number of nodal lines in the low-energy fanlike interference structures \cite{Shvetsov2016}.

With various combinations of the above methods, including methods to prepare initial conditions (SPA, SPANE, and ADK) and methods to weight the electron trajectories (CTMC, QTMC, and SCTS), Ma \textit{et al.}\ \cite{Ma2021} systematically studied the influence of nonadiabatic and quantum effects on the attoclock signal. It has been shown that both nonadiabatic and quantum effects play a crucial role in shaping the momentum distribution within the attoclock protocol. To obtain a consistent interpretation of the attoclock signal, therefore, both effects need to be carefully accounted for.

Although trajectory-based methods are effective tools for providing clear physical origins of Coulomb-induced influence on attoclock signals, they do not directly answer the question whether tunneling takes time. For this reason, a method including both quantum tunneling dynamics and clear physical pictures is still in need.

\subsection{Backpropagation method}

\begin{figure*}[t]
\centering
\includegraphics[width=\textwidth]{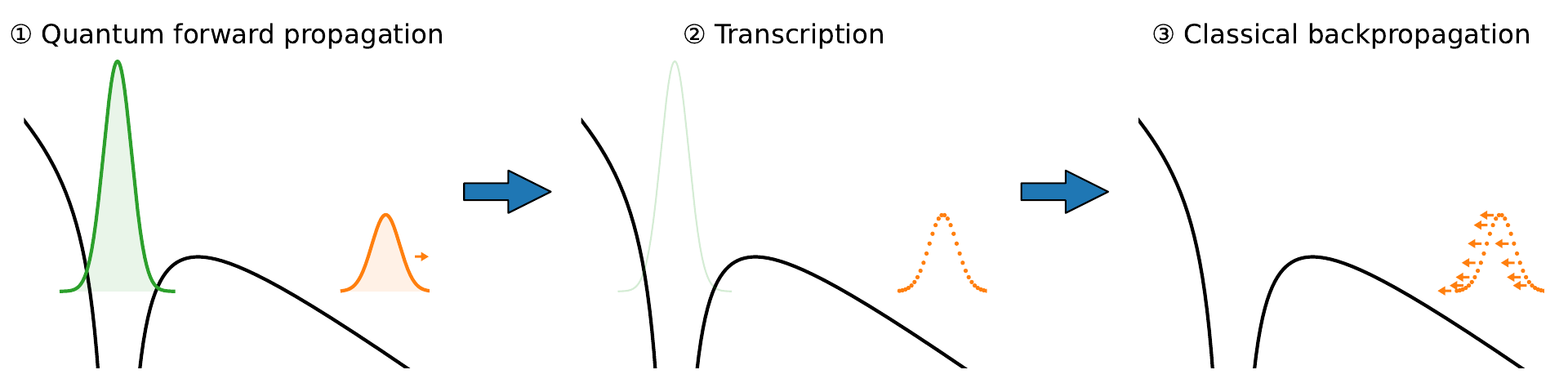}
\caption{Concept of the backpropagation method.}
\label{fig:backprop}
\end{figure*}

Tunneling is without any doubt a purely quantum process, which has been frequently used as an illustration of the quantum nature of a microscopic object, and it is fundamentally nonlocal. When one asks about the tunneling characteristics, such as the ``position'', ``momentum'', and ``time'' corresponding to the tunnel exit, however, one introduces local elements. How can one answer the question regarding the tunneling exit time and position for a quantum particle? Is there a way to fully incorporate quantum tunneling while at the same time retaining clear physical pictures and local properties such that one may extract the tunneling exit characteristics? In 2016, Ni \textit{et al.}\ proposed a novel backpropagation method \cite{Ni2016,Ni2018a,Ni2018b}, which combines quantum forward propagation and classical backpropagation, providing thereby a unique perspective on the problem of tunneling time delay. It has been extensively applied to investigate various aspects of the tunneling process, including characterization of the tunnel exit \cite{Ni2018b,Klaiber2022}, retrieval of tunneling \cite{Wang2017,Zhang2017} and deformation dynamics \cite{Liu2018} of atomic $p$ orbitals, resolution of rescattering time \cite{Kim2021}, and study of subcycle nondipole transfer of linear momentum \cite{Ni2020}.

\subsubsection{Implmentation of backpropagation}

The backpropagation method is a hybrid quantum-classical method. The unusual combination of a quantum forward propagation and a classical backpropagation brings about new possibilities of extracting classical local features from nonlocal quantum processes without approximating the actual quantum dynamical evolution. In the case of treating tunneling problems, it enables the retrieval of tunneling exit information without approximating the tunneling process itself, whose accurate description is a prerequisite to define the tunnel exit.

As shown in Fig.~\ref{fig:backprop}, the backpropagation method consists of three steps: (1) quantum forward propagation, (2) transcription of eventually freed quantum wave packet into classical trajectories, and (3) classical backpropagation of the trajectories backward in time until reaching the tunnel exit.

In the first step, a fully quantum mechanical evolution is carried out starting from the ground initial state of the target. Thereby, an electron wave packet is freed without invoking any assumptions or approximations. Note that the ionizing quantum wave packet cannot be separated from the full wave function during the under-barrier tunneling process due to quantum nonlocality. When the laser ceases, however, the freed portion and the bound portion of the total wave function is detached spatially, enabling the isolation asymptotically.

In the second step, the eventually freed portion of the quantum wave packet is picked spatially and transcribed into classical trajectories at a time $t_f$ after the laser pulse. At each spatial grid point $\bm{r}$ within the freed wave packet, a classical trajectory is launched, which possesses a weight equal to the probability $|\psi(\bm{r},t_f)|^2$ of the wave function at that grid point $\bm{r}$. The momentum $\bm{k}$ of the classical trajectory at $\bm{r}$ can be obtained from the local-momentum method \cite{Feuerstein2003,Wang2013,Wang2018,Xu2021}:
\begin{equation}
\bm{k}(\bm{r},t_f) = \bm{\nabla}S(\bm{r},t_f),
\label{eq:localmomentum}
\end{equation}
where $S(\bm{r},t_f)=\arg\{\psi(\bm{r},t_f)\}$ is the phase of the quantum wave packet at grid point $\bm{r}$. Notably, the information of both the amplitude and the phase of the quantum wave packet is retained in the constructed ensemble of classical trajectories. While the amplitude is contained in the weight sampling, the phase is retained in the momentum of the trajectories. The local momentum as defined in Eq.~\eqref{eq:localmomentum} can be rigorously derived from the Wigner function in the saddle-point approximation in the limit $\hbar\rightarrow0$ \cite{Jin2003}, corresponding to the peak of the momentum distribution at $\bm{r}$.

In the third step, the trajectories are propagated backward in time following the Newton's equation of motion
\begin{equation}
\ddot{\bm{r}}=-\bm{\nabla}V(r)-\bm{F}(\bm{r},t)
\end{equation}
until a criterion defining the tunnel exit is met. It is worthwhile noting that there could be multiple solution when a specific criterion is met, and the true solution is the one that is closest to the nucleus.

An alternative way to implement the backpropagation method is to place a sphere of homogeneously distributed virtual detectors around the nucleus \cite{Feuerstein2003,Wang2013,Wang2018,Xu2021}. Whenever some quantum flux hits a specific virtual detector at $\bm{r}_d$, a classical trajectory is launched with momentum
\begin{equation}
\bm{k}(\bm{r}_d,t)=\bm{\nabla}S(\bm{r}_d,t),
\end{equation}
which is similar to Eq.~\eqref{eq:localmomentum} but assessed at specific grid points $\bm{r}_d$ where the virtual detectors locate for different time instances $t$ during the entire laser pulse. The weight associated with a specific trajectory is $\bm{j}(\bm{r}_d,t)\cdot\hat{\bm{n}}(\bm{k}_d)$, where $\hat{\bm{n}}(\bm{k}_d)$ is the norm vector at detector $\bm{r}_d$. Negative weight corresponds to ingoing flux which is usually very small compared to outgoing flux, and can be neglected. The detector sphere should be placed sufficiently far away from the nucleus to ensure a small ingoing-to-outgoing ratio of the flux. For three-dimensional computations, this alternative implementation is typically less resource demanding. In addition, it also allows the employment of long laser pulses since intercycle interference that contaminates the phase distribution needed for backpropagation has not emerged yet when the quantum flux reaches the detector sphere.

\subsubsection{Tunneling criteria}

Formally, the classical trajectories propagate in the phase space spanned by the coordinate and momentum of the particles. As the trajectories are propagated backward in time, a projector in the phase space is necessary to define the tunnel exit, where backpropagation stops. The condition to stop the backpropagation, or the tunneling criterion, can be defined in the coordinate space or in the momentum space, or a combination of them as well as time. As the three basic steps of backpropagation are the same, the backpropagation method provides a means to test different tunneling criteria on the same footing, in an aim to find the most suitable definition of the tunnel exit \cite{Ni2018b}, thereby providing a consistent basis for characterizing tunneling ionization including tunneling delay. The tunneling criteria can generally be divided into three categories: velocity-based criteria, position-based criteria, and energy-based criteria.

{\it Velocity-based tunneling criteria.} As a straightforward extension from the one-dimensional tunneling scenario, the momentum parallel to the instantaneous electric field direction $k_\parallel$ at the tunnel exit has often been set to be zero, which is the velocity criterion of tunneling
\begin{equation}
k_\parallel = -\bm{k}\cdot\hat{\bm{F}} = 0.
\label{eq:velocityCriterion}
\end{equation}
It can be routinely derived from the SPANE method, as illustrated in Eq.~\eqref{eq:imagSPANE}. If one is to carry out the nonadiabatic expansion to higher orders, an even more accurate elliptical velocity criterion results:
\begin{equation}
k_\varepsilon = -\bm{k}\cdot\hat{\bm{F}} - \frac{3\bm{F}\cdot\dot{\bm{F}}-\bm{k}\cdot\ddot{\bm{F}}}{6F} \frac{k^2+2I_p}{F^2-\bm{k}\cdot\dot{\bm{F}}} = 0.
\label{eq:ellipticalVelocityCriterion}
\end{equation}
One could also take the Coulomb potential into account and define the parabolic velocity criterion
\begin{equation}
k_\eta = \frac{d}{dt}\left(r-\hat{\bm{F}}\cdot\bm{r}\right) = 0,
\end{equation}
define tunneling to occur at the turning point of the backpropagating trajectories
\begin{equation}
k_r = \bm{k}\cdot\hat{\bm{r}} = 0,
\end{equation}
or define tunneling by the local minimum in the speed
\begin{equation}
\dot{k} = \frac{d}{dt}|\bm{k}| = 0.
\end{equation}

It has been shown that all these velocity-based tunneling criteria result in similar characteristics of the tunnel exit, leading to a tunneling time delay close to zero. Notably, these velocity-based criteria do not make assumptions regarding to the instantaneous tunneling energy, which is important to ensure consistency.

{\it Position-based tunneling criteria.} Another category of commonly used tunneling criteria is based on the position of the tunnel exit. These criteria include the position criterion
\begin{equation}
r = \frac{I_p+\sqrt{I_p^2-4\beta F}}{2F}
\end{equation}
with $\beta=1-\sqrt{2I_p}/2$ the separation constant in parabolic coordinates, the Stark position criterion
\begin{equation}
r = \frac{I_p^\mathrm{Stark}+\sqrt{\left(I_p^\mathrm{Stark}\right)^2-4\beta^\mathrm{Stark}F}}{2F}
\end{equation}
incorporating Stark shift to the ionization potential where $\beta^\mathrm{Stark}=1-\sqrt{2I_p^\mathrm{Stark}}/2$ with $I_p^\mathrm{Stark}=I_p+\frac{1}{2}\alpha F^2$ and $\alpha$ the polarizability of the initial state, and the dynamic position criterion
\begin{equation}
r = \frac{-E+\sqrt{E^2-4\beta^\mathrm{inst}F}}{2F}
\end{equation}
accounting for the instantaneous binding energy $-E$ with $\beta^\mathrm{inst}=1-\sqrt{-2E}/2$.

The position-based tunneling criteria typically lead to positive tunneling time delays. However, these criteria suffer from severe problems, which will be detailed later.

{\it Energy-based tunneling criteria.} There have also been attempts to define the tunnel exit based on the total energy of the electrons, including the static energy criterion
\begin{equation}
E = -I_p
\end{equation}
specifying that the tunneling energy is the value of the static ground-state energy, and the Stark energy criterion
\begin{equation}
E = -I_p^\mathrm{Stark}
\end{equation}
incorporating Stark shift to the ground-state energy. These energy-based tunneling criteria suffer from severe problems similar to the case of position-based criteria.

\subsubsection{Optimal tunneling criteria}

The classical backpropagation method provides a common ground to investigate and compare different definitions of the tunneling criteria on the same footing, thereby providing a consistent physical description of tunneling ionization.

In the scenario of one-dimensional static tunneling, velocity-based criteria and position-based criteria are fully equivalent, leading to identical characteristics of the tunnel exit. Energy-based criteria are on the other hand fulfilled at all times, making them completely unsuitable as conditions to define tunneling.

When nonadiabaticity comes into play, e.g., as in tunneling ionization in intense laser fields, velocity-based and position-based criteria are no longer equivalent. The reason is that, for position-based criteria to apply, an assumption regarding the tunneling energy needs to be made to obtain the tunneling position in the first place. However, if one backpropagates the classical trajectories till the designated positions, the energy turns out different from that assumed, leading to inconsistencies. In addition, a large fraction of trajectories cannot reach these designated locations. This means that, no combinations of initial conditions would reproduce the correct asymptotic momentum for these trajectories. On the other hand, velocity-based criteria do not make any assumptions regarding the instantaneous tunneling energy, making them consistent definitions of tunneling \cite{Ni2018b}.

\begin{figure}[tb]
\centering
\includegraphics[width=\columnwidth]{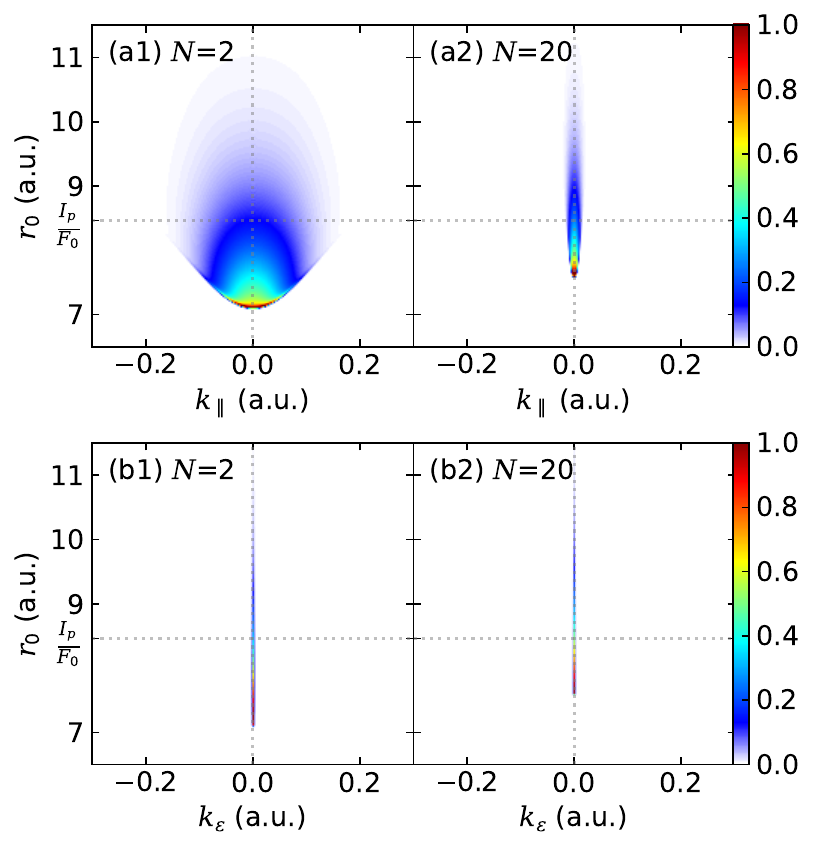}
\caption{Correlated distribution of the (a) parallel tunneling momentum $k_\parallel$ [Eq.~\eqref{eq:velocityCriterion}] or (b) elliptical tunneling momentum $k_\varepsilon$ [Eq.~\eqref{eq:ellipticalVelocityCriterion}] and the tunneling exit radius $r_0$ of the helium atom in a 800 nm circularly polarized laser pulse with a peak intensity of $8\times10^{14}$ W/cm$^2$ and a total number of cycles (1) $N=2$ and (2) $N=20$ calculated with SPA.}
\label{fig:kpr0}
\end{figure}

To further illustrate the applicability of the velocity-based and position-based criteria, we show in Fig.~\ref{fig:kpr0}(a) the correlated distribution of the parallel tunneling momentum $k_\parallel$ [Eq.~\eqref{eq:velocityCriterion}] and the tunneling exit radius $r_0$ of the helium atom in a 800 nm circularly polarized laser pulse with a peak intensity of $8\times10^{14}$ W/cm$^2$ and a total number of cycles $N=2$ (left column) or $N=20$ (right column) calculated with SPA. The correlated distribution can be used as a straightforward guide to judge how good a particular tunneling criterion is. As obvious from the figure, the parallel tunneling momentum $k_\parallel$ has generally a narrow profile of distribution around 0, even under ultrashort laser pulses with 2 total number of cycles. The tunneling exit radius $r_0$, on the other hand, concentrates sharply and much closer to the nucleus than the expected location around $r_0=I_p/F_0$ due to nonadiabatic energy absorption during the under-barrier motion. Clearly, the velocity criterion is much superior than the position criterion. A further improved velocity-based criterion, i.e., the elliptical velocity criterion $k_\varepsilon=0$ [Eq.~\eqref{eq:ellipticalVelocityCriterion}] demonstrates an even better performance than the velocity criterion $k_\parallel=0$ [Eq.~\eqref{eq:velocityCriterion}], as depicted in Fig.~\ref{fig:kpr0}(b). There, the $k_\varepsilon$ has an extremely narrow profile of distribution around 0, making it a superior criterion to define tunneling. We note that, for all velocity-based criteria, a near-zero tunneling time delay results in the tunneling regime, providing the evidence that tunneling ionization is essentially instantaneous.

In brief, velocity-based criteria that take full account of nonadiabaticity should be applied to define tunneling, which leads to a close-to-instantaneous tunneling. A finite tunneling delay could be a misinterpretation that results from the application of position-based criteria that suffer from inconsistencies and difficulties.

\subsubsection{Depletion correction}

The backpropagation method is capable of extracting the fully differential tunneling exit characteristics including the tunneling ionization time $t$, giving the tunneling ionization rate $P(t)$. From this tunneling ionization rate, one can correct for the effect of depletion
\begin{equation}
\tilde{P}(t) = \frac{P(t)}{1-\int_{-\infty}^t dt'P(t')},
\end{equation}
which is free from the depletion effect, subsequently leading to the mean tunneling ionization time in the absence of depletion \cite{Ni2018a}
\begin{equation}
\langle\tilde{t}\rangle = \int_{-\infty}^{\infty}dt t\tilde{P}(t) = \int_{-\infty}^{\infty}dt\frac{tP(t)}{1-\int_{-\infty}^t dt'P(t')}.
\end{equation}

\subsubsection{Nontunneled fraction}

Within the laser parameter range where ionization occurs in the tunneling regime, most classical trajectories are expected to stop at the tunnel exit defined by the tunneling criteria, i.e., the nontunneled fraction should be very low. This is the case with velocity-based criteria, typically below the level of 0.01\% \cite{Ni2016,Ni2018a}. However, the nontunneled fraction could be as high as 30\% for position-based criteria or 90\% for energy-based criteria \cite{Ni2018b}, making them unsuitable candidates to define tunneling.

\begin{figure}[b]
\centering
\includegraphics[width=\columnwidth]{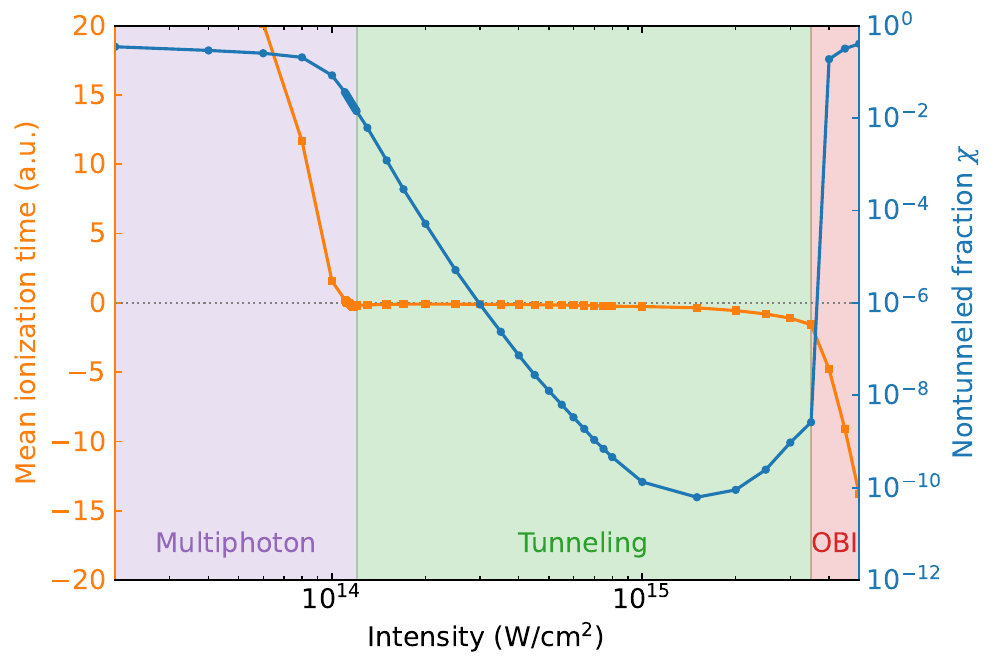}
\caption{Intensity dependence of the nontunneled fraction of the ionization probability (right ordinate) and the mean tunneling ionization time (left ordinate, corrected for depletion) of the helium atom in a two-cycle 1000 nm circularly polarized laser pulse using the velocity tunneling criterion \cite{Ni2018a}.}
\label{fig:fraction}
\end{figure}

In addition, the nontunneled fraction may serve as an indicator of the regime the ionization occurs in. Shown in Fig.~\ref{fig:fraction} is the intensity dependence of the nontunneled fraction of the ionization probability (right ordinate) and the mean tunneling ionization time (left ordinate, corrected for depletion) of the helium atom in a two-cycle 1000 nm circularly polarized laser pulse using the velocity tunneling criterion \cite{Ni2018a}. When the intensity is relatively low, i.e., under $1.2\times10^{14}$ W/cm$^2$, the ionization occurs in the multiphoton regime, the nontunneled fraction is, not surprisingly, high. Correspondingly, the mean ionization time under this scenario becomes very large and indeed loses its meaning as tunneling exit time. Conversely, when the intensity is very high, i.e., above $3.5\times10^{15}$ W/cm$^2$, the ionization occurs in the over-barrier regime, leading to over-barrier ionization (OBI), the nontunneled fraction also reaches a very high value and the ionization time reaches a big negative value even after correction for depletion. Within the intensity range between $1.2\times10^{14}$ and $3.5\times10^{15}$ W/cm$^2$, the ionization occurs in the tunneling regime, where the nontunneled fraction is very low and the corresponding mean tunneling ionization time is very close to zero after correction for depletion. The small deviation from zero roots in the under-barrier recollision process \cite{Klaiber2022,Klaiber2023}. Interestingly, the transition between multiphoton and tunneling ionization is gradual while the transition from tunneling to over-barrier ionization is quite abrupt.

\subsection{Recent advances}

In order to further clarify whether there is a tunneling ionization time delay, revisions to the attoclock techniques have been proposed recently. In 2019, Han {\it et al.}\ \cite{Han2019} utilized an improved attoclock technique attempting to unify the two contradictory views of the tunneling time delay. The improved attoclock is composed of a strong circularly polarized 800 nm laser field and a perturbative linearly polarized 400 nm laser field. Due to the existence of the weak linear field, the symmetry of the light field is broken, so the momentum distribution is no longer an isotropic ring. Rather, ionization peaks along the polarization direction of the weak 400 nm field. Similar to the attoclock offset angle, this direction can be modeled under the same framework. The advantage of this improved attoclock is that it is experimentally possible to use circularly polarized laser field with longer pulse durations and does not require precise control over the carrier envelope phase and ellipticity. Based on the improved attoclock technique, Han {\it et al.}\ verified that the offset angle is caused by the Coulomb potential using SFA, Coulomb-corrected strong-field approximation (CCSFA) and TDSE simulations. Further, the time delay corresponding to the most probable Wigner trajectory \cite{Camus2017} can be obtained within the same framework of SFA, resulting in a positive finite tunneling time delay and a nonzero longitudinal momentum at the tunnel exit. As a matter fact, the reason the Wigner trajectory leads to a positive delay is that the full wave function near the tunnel exit is assessed, instead of the eventually freed portion of the wave function. This situation is similar to that found using Bohmian trajectories, where the trajectories are constructed on top of the pilot wave composed of the entire wave function near the tunnel exit \cite{Douguet2018}.

The improved attoclock has been further extended by Yu {\it et al.}\ \cite{Yu2022}, attempting to tackle the problem of tunneling ionization time independent of theoretical models. To this end, a linearly polarized weak 400 nm pulse is superimposed on the elliptically polarized strong 800 nm laser field. The total electric field strength at a certain time oscillates with the relative phase $\Phi$ between the two pulses. Correspondingly, the ionization yield at a certain offset angle also oscillates with the relative phase $\Phi$, constructing a phase-of-phase spectroscopy. Fitting the yield as a function of $\Phi$ shows that the yield peak is synchronous with the electric field peak. The authors thereby claim that tunneling ionization is instantaneous within the experimental precision. However, later works \cite{Klaiber2023} claim that such an experimental protocol involves only a parametric dependence of the ionization yield with the field strength and does not involve dynamical tunneling, and therefore cannot be used as an evidence of instantaneous tunneling.

Another variant of the phase-of-phase attoclock involves a strong elliptically polarized 800 nm laser field and a weak co-rotating circularly polarized 400 nm laser pulse \cite{Han2021}. The weak 400 nm pulse breaks the symmetry in the elliptical pulse such that the two lobes of the attoclock momentum distribution are no longer symmetric. As the relative phase between the two pulses varies, the relative strength of the two lobes also varies, providing thereby richer information than the common single-color attoclock, which enables the retrieval of not only the real tunneling time but also the imaginary tunneling time and the quantum phase accumulated under the tunneling barrier.

By combining two-color counter-rotating bicircular polarized laser pulses, the pulse profile can even be tailored such that ``it approximates linear polarization three times per optical cycle of the fundamental component, while providing a time-to-momentum mapping similar to the attoclock'' \cite{Eicke2019,Eicke2020}, thereby enabling retrieval of tunneling time delay in a quasilinear laser geometry.

While the attoclock signal has been typically interpreted with the help of trajectories, a trajectory-free formulation has been proposed by Eicke {\it et al.}\ \cite{Eicke2018} by assessing the saddle-point time of the Dyson integral of the full dynamical evolution governed by the TDSE. With this trajectory-free approach, a mapping between the final momentum and the ionization time can be established. It has been concluded from the study of the full quantum dynamical evolution that tunneling ionization is instantaneous.

The attoclock technique can be applied not only to atoms but also to molecules. Serov {\it et al.}\ \cite{Serov2019} showed that the attoclock deflection angle for the hydrogen atom and the hydrogen molecule is similar, although the width of the attoclock angular distribution depends on the molecular orientation. Quan {\it et al.}\ \cite{Quan2019} proposed a molecular hydrogen attoclock based on the ionic fragments of H$_2^+$ and H$^+$, demonstrating nearly negligible tunneling time delay for the dissociative ionization of the hydrogen molecule. For diatomic molecules with large internuclear distances, however, interesting ionization dynamics appears. Tong {\it et al.}\ \cite{Tong2022} demonstrated that, after single ionization from the Kr side in the Ar-Kr dimer, the second electron may either tunnel out directly from the Ar side or get trapped by the binary potential before its release, inducing thereby a transient trapping time delay of about 3.5 fs. In addition, Guo {\it et al.}\ \cite{Guo2021} and Trabert {\it et al.}\ \cite{Trabert2021NC} have attempted to connect the anisotropic tunneling of the hydrogen molecule to the Wigner time delay.

\subsection{Towards a consensus of tunneling time delay}

The existence of tunneling time delay has been a long-debated problem since the birth of quantum mechanics, which has resurrected with the recent introduction of the attoclock technique. While interpreting the attoclock signal, a few attempts have been made towards a consensus of the issue of tunneling time delay. In 2018, Ni {\it et al.}\ \cite{Ni2018a,Ni2018b} utilized the backpropagation method as a common ground to investigate the attoclock signal. It has been shown that when nonadiabatic tunneling effects are fully and consistent taken account of using the velocity-based tunneling criteria, the tunneling time delay is close to zero. While if position-based criteria are used, which only partially accounts for nonadiabaticity and suffers from inconsistencies and difficulties, the tunneling time delay is a positive value. Therefore, a finite tunneling time delay could be a misinterpretation of the attoclock where tunneling is not defined in an appropriate way.

In 2019, Han {\it et al.}\ \cite{Han2019} attempted to bring together the two controversial views of tunneling time delay using the same framework of SFA. It was found that the attoclock offset angle arises purely from the Coulomb interaction. Assessing the attoclock signal, therefore, an instantaneous tunneling can be concluded. However, if one studies the Wigner trajectory \cite{Camus2017} corresponding to the optimal probability during the tunneling process, a positive tunneling delay results.

In 2022, Klaiber {\it et al.}\ \cite{Klaiber2022,Klaiber2023,Bakucz2021} explicitly sorted presently studied tunneling ionization time delays into three categories: the asymptotic time delay (ATD) as deduced from the asymptotic momentum distribution of the attoclock signal, the exit time delay (ETD) as following from a \textit{Gedankenexperiment} assessing the wave packet near the tunnel exit, and an initiation time delay (ITD) describing the initiation of the tunneling wave packet. It was concluded that the ATD is zero in the deep tunneling regime, and could become slightly negative due to under-barrier recollision when the laser intensity increases to be close to OBI. On the other hand, the ETD is positive, such as that found with the Wigner \cite{Camus2017} and Bohmian \cite{Douguet2018} trajectory methods, since they are both based on the entire wave packet near the tunnel exit instead of the eventually freed portion of the quantum wave function. In addition, ITD, such as that found in the two-color attoclock setup \cite{Yu2022}, is zero, which, however, represents only a parametric dependence of the yield on the electric field strength and does not involve dynamical tunneling.

\begin{figure*}[tb]
\centering
\includegraphics[width=\textwidth]{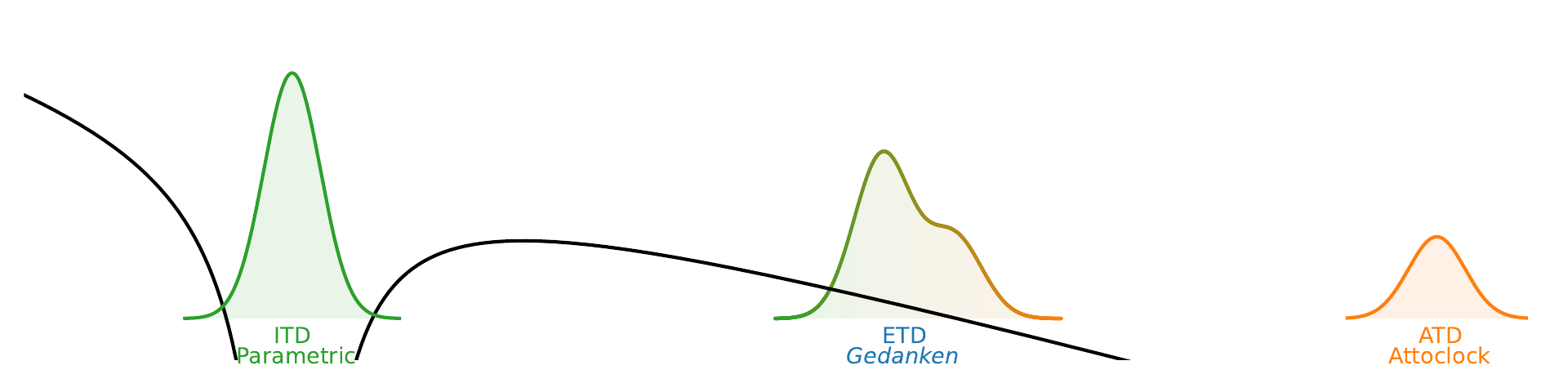}
\begin{equation*}
\text{Tunneling Time Delay}
\begin{cases}
\text{Asymptotic Time Delay (ATD)} &
\begin{cases}
\text{Nonadiabaticity consistently included}\quad
\begin{cases}
\text{Deep Tunneling} & t_\mathrm{tun}^\mathrm{ATD}=0 \\[6pt]
\text{Near Over-Barrier} & t_\mathrm{tun}^\mathrm{ATD}\lesssim0
\end{cases}\\[16pt]
\text{Nonadiabaticity not fully or consistently included}\quad t_\mathrm{tun}^\mathrm{ATD}>0
\end{cases}\\[25pt]
\text{Exit Time Delay (ETD)} & t_\mathrm{tun}^\mathrm{ETD}>0 \\[6pt]
\text{Initiation Time Delay (ITD)} & t_\mathrm{tun}^\mathrm{ITD}=0
\end{cases}
\end{equation*}
\caption{Categorization of tunneling ionization time delays.}
\label{fig:category}
\end{figure*}

The present scope and consensus of the tunneling time delay problem can be summarized as in Fig.~\ref{fig:category}. We stress that, to assess the attoclock experimental signal, the ATD should be used as it corresponds to the information extracted from the asymptotically measurable momentum distribution.

\section{Summary and perspective}
\label{sec:summary}

In summary, we have overviewed in this Topic Review the attosecond interrogation techniques and the associated attosecond electronic dynamical processes these techniques have enabled to time resolve. The interrogation techniques include the attosecond streak camera, the RABBITT, and the attoclock techniques. The attosecond electronic processes include photoionization and tunneling ionization from atomic and molecules targets. The basic formulations of the interrogation techniques and physical processes are placed under a unified theoretical framework, providing a consistent physical description. In particular, the novel hybrid quantum-classical backpropagation method has been overviewed, which has enabled complete and consistent characterization of the tunneling ionization process. We have also included a few lately and heatedly debated topics and overviewed recent advances towards a consensus.

In addition to the advancements in exploring quantum dynamics at ever faster rates and on ever shorter timescales, interdisciplinary integration with other fields hold promising prospects, leading to ultrafast chemistry \cite{Li2022,Zhou2023,Mi2023}, ultrafast nanophysics \cite{Ciappina2017,Seiffert2022}, and ultrafast quantum information science \cite{Lewenstein2021,Lewenstein2022,Gorlach2020,Gorlach2023,Tzur2023,Fang2023}, offering fertile ground for innovative research. Continued exploration in the realm of ultrafast science promises an ever broader scope, contributing to the expansion of human knowledge on an ever increasing scale.


\section*{Acknowledgments}

We would like to thank Mingyu Zhu for careful proofreading. This work was supported by the National Natural Science Foundation of China (Grant Nos.\ 92150105, 11834004, 12227807, and 12241407), and the Science and Technology Commission of Shanghai Municipality (Grant No.\ 21ZR1420100).

\end{document}